\documentclass[journal=ancac3,manuscript=article]{achemso}
\usepackage[version=3]{mhchem} \usepackage{amssymb,amsfonts,amsmath}
\usepackage{graphicx}
\usepackage[normalem]{ulem}
\usepackage{float}
\usepackage{color}
\usepackage{hyperref}\hypersetup{backref,colorlinks=true}
\numberwithin{equation}{section}

\newcommand{\uvec}[1]{\mathbf{\hat{{#1}}}}
\renewcommand{\vec}[1]{\mathbf{#1}}
\newcommand{\kbT}{k_{\scriptscriptstyle\rm B}T}
\SectionNumbersOn
\AtBeginDocument{\singlespacing}

\author{Sophia Jordens}
\affiliation{ETH Zurich, Department of Health Sciences \& Technology, Laboratory of Food \& Soft Materials, 8092 Zurich, Switzerland}
\author{Emily E. Riley}
\affiliation{University of Leeds, School of Physics \& Astronomy, Soft Matter Physics Group, Leeds LS2 9JT, United Kingdom}
\altaffiliation{Current address: Cambridge University, Department of Applied Mathematics and Theoretical Physics, Cambridge CB3 0WA, United Kingdom}
\author{Ivan Usov}
\affiliation{ETH Zurich, Department of Health Sciences \& Technology, Laboratory of Food \& Soft Materials, 8092 Zurich, Switzerland}
\author{Lucio Isa}
\affiliation{ETH Zurich, Department of Materials, Laboratory for Surface Science \& Technology, 8093 Zurich, Switzerland}
\altaffiliation{Current address: ETH Zurich, Department of Materials, Laboratory for Interfaces, Soft Matter \& Assembly, 8093 Zurich, Switzerland}
\author{Peter D. Olmsted}
\email{pdo7@georgetown.edu}
\affiliation{University of Leeds, School of Physics \& Astronomy, Soft Matter Physics Group, Leeds LS2 9JT, United Kingdom}
\altaffiliation{Current Address: Georgetown University, Department of Physics and Institute for Soft Matter Synthesis \& Metrology, Washington DC 20057, USA}
\author{Raffaele Mezzenga}
\email{raffaele.mezzenga@hest.ethz.ch}
\affiliation{ETH Zurich, Department of Health Sciences \& Technology, Laboratory of Food \& Soft Materials, 8092 Zurich, Switzerland}
\keywords{$\beta$-lactoglobulin, amyloid fibrils, biopolymers, interfaces, bending, statistical analysis, atomic force microscopy}

\title{Adsorption at Liquid Interfaces Induces Amyloid Fibril Bending and Ring Formation}

\begin{document}
\setstretch{0.95}

\begin{center}{\includegraphics[width=0.5\textwidth]{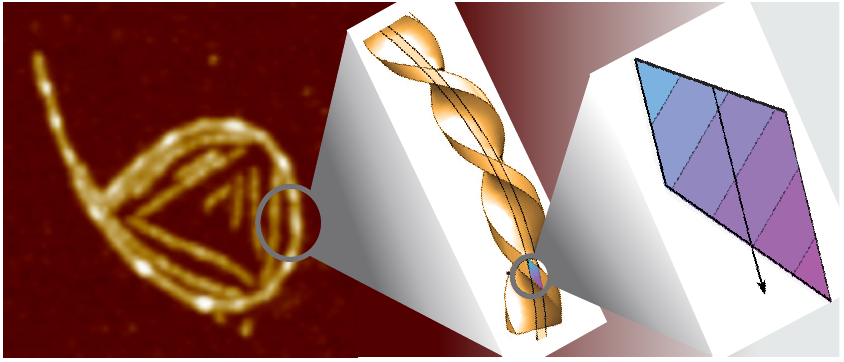}}\end{center}
This document is the unedited Author's version of a Submitted Work that
was subsequently accepted for publication in ACS \copyright
American Chemical Society after peer review. To access the final edited
and published work see \href{http://pubs.acs.org/doi/abs/10.1021/nn504249x}{http://pubs.acs.org/doi/abs/10.1021/nn504249x}.
\newpage

\begin{abstract}
Protein fibril accumulation at interfaces is an important step in many physiological processes and neurodegenerative diseases as well as in designing materials. Here we show, using $\beta$-lactoglobulin fibrils as a model, that semiflexible fibrils exposed to a surface do not possess the Gaussian distribution of curvatures characteristic for wormlike chains, but instead exhibit a spontaneous curvature, which can even lead to ring-like conformations. The long-lived presence of such rings is confirmed by atomic force microscopy, cryogenic scanning electron microscopy and passive probe particle tracking at air- and oil-water interfaces. We reason that this spontaneous curvature is governed by structural characteristics on the molecular level and is to be expected when a chiral and polar fibril is placed in an inhomogeneous environment such as an interface. By testing $\beta$-lactoglobulin fibrils with varying average thicknesses, we conclude that fibril thickness plays a determining role in the propensity to form rings.
\end{abstract}

\section{Introduction}
Polymers exposed to an unfavorable environment can collapse
or change shape in order to minimize surface energy
\cite{deGennes,Doi,Pereira}. Examples of unfavorable environments
include a poor solvent or a hydrophilic-hydrophobic interface like the
one between water and either air or oil. Examples of conformations driven by such energy minimization are
rings, loops, coils, spools, tori/toroids, hairpins or tennis rackets
\cite{Cohenmorph}. In filaments comprising aggregated proteins or
peptides, ring formation falls into two main classes: fully annealed
rings occasionally observed as intermediate states during protein
fibrillation, like in apolipoprotein C-II \cite{Hatters} and
A$\beta_{1-42}$ \cite{Mustata}; or ring formation in actively driven
systems, where the energy required for filament bending is provided by GTP or ATP \cite{Paez,Tang,Kabir,Sumino}. 
Insulin has been shown to
form open-ring shaped fibrils when pressure was applied during
fibrillation \cite{Jansen}, which was explained by an anisotropic
distribution of void volumes in fibrils and therefore aggregation into
bent fibrils.

We study amyloid fibrils, which are linear supramolecular assemblies
of proteins/peptides that, despite a large diversity in possible
peptide sequences, show remarkable structural homogeneity. Peptides
form $\beta$-sheets that stack, often with chiral registry, to form a
filament whose main axis is perpendicular to the $\beta$-strands
\cite{Dobson, Eichner}. Fully formed fibrils can consist of one or,
more commonly, multiple filaments, assembled into twisted ribbons with
a twist pitch determined by the number of filaments in the
fibrils \cite{Adamcik}. Their high aspect ratio (diameter usually less
than $10$~nm, total contour length up to several $\mu$m) leads to liquid
crystalline phases in both three (3D) \cite{Jung} and two dimensions
(2D) \cite{IsaSM,Jordens}. Amyloid fibrils were initially studied due
to their involvement in many different degenerative diseases such as
diabetes II or Parkinson's disease \cite{DobsonNature}. However,
protein fibrils have recently experienced a surge of interest in
potential applications in materials \cite{Mankar}, and functional
roles have been identified in biological processes such as hormone
storage \cite{Riek}, emphasizing the importance of understanding their
structure and properties in 2D.

Here, we present experimental evidence for the development of
\textit{curved} fibrils at interfaces.  Semiflexible
$\beta$-lactoglobulin fibrils are found to undergo a shape change
and passively form open rings upon adsorption to an interface
(liquid-liquid or liquid-air). We show that this cannot be described
by a simple bending modulus; this bending can instead be understood in terms
of a \textit{spontaneous curvature} induced on symmetry grounds by the
chiral and polar nature of the fibril, when interacting with the
heterogeneous environment provided by an interface. A comparison of
different fibril batches of the same protein shows that the
probability of forming rings depends on the average fibril thickness,
with batches of thicker fibrils not forming loops. These results imply
that flexible non-symmetric bodies embedded in heterogeneous media
\textemdash ~such as the physiological environment \textemdash ~can be
expected to deform, bend, and twist, depending on the specific surface
interaction with the environment. For example, concentration gradients
of ions or pH could enhance shape changes necessary for locomotion in
flexible nanoswimmers \cite{Sengupta,Keaveny}, or be used to promote
or control self-assembly through shape changes. One could even
envision high surface to volume materials such as bicontinuous phases
with large length scales being used to process large amounts of
flexible shape changers.

\section{Results and discussion}
\subsection{Morphology}
\begin{figure*}[h!]
\begin{center}
{\includegraphics[width=0.9\textwidth]{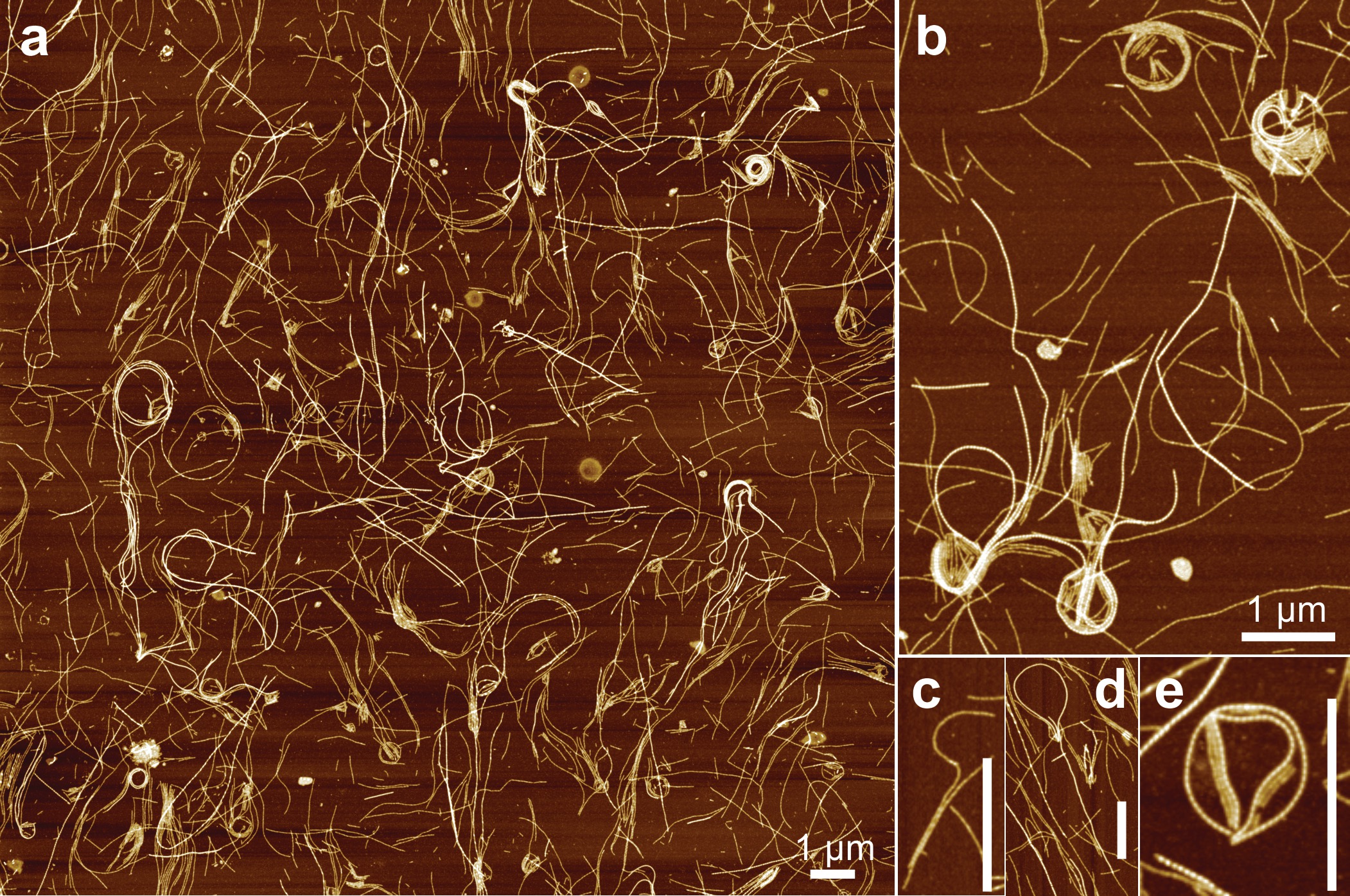}}
\end{center}
\caption{ $\beta$-lactoglobulin fibrils form rings and loops at the air-water interface. AFM images of $\beta$-lactoglobulin fibrils at the air-water
  interface after horizontal transfer onto mica using the
  Langmuir-Schaefer method. The initial fibril concentration in the
  bulk was $c_{\text{init}}=0.001\%$~w/w and the waiting time before transfer
  $t=60$ minutes. The scale bars in {\bf c}, {\bf d}, and {\bf e} correspond to
  $1$~$\mu$m.}
\label{fig:AFM}
\end{figure*}

\begin{figure*}
\begin{center}
{\includegraphics[width=1\textwidth]{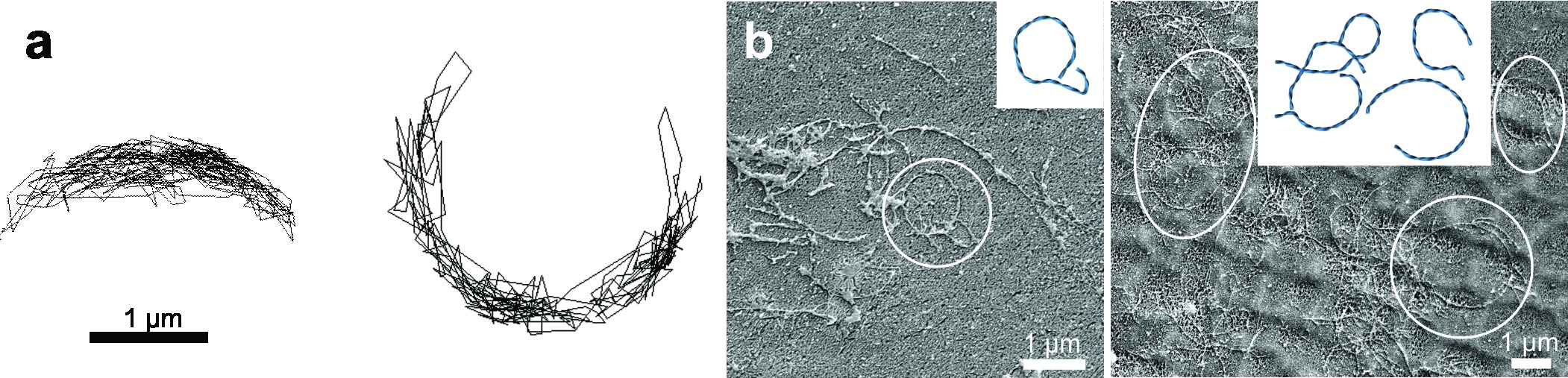}}
\end{center}
\caption{ Fibril rings at an oil-water interface. ({\bf a}) Trajectories of
  two different silica tracer particles with diameter
  $\approx{774}$~nm moving around circular obstacles during a passive
  probe particle tracking experiment of a $c_{\text{init}}=0.005\%$~w/w
  fibril suspension. The trajectories were isolated $t=24$~and~$37$
  minutes after interface creation, respectively. The scale bar
  applies to both trajectories. ({\bf b}) Cryo-SEM images of the surface of
  a $c_{\text{init}}=0.001\%$~w/w fibril suspension $t=60$ minutes after
  medium chain triglycerides (MCT)-water interface creation. The oil
  phase had been removed by freeze-fracture prior to imaging. Insets
  show sketches of the rings indicated with circles in the main
  images.}
\label{fig:PTSEM}
\end{figure*}

When imaging the air-water interfacial fibril layer by AFM using a
modified Langmuir-Schaefer horizontal transfer technique (see
Materials and Methods) to resolve 2D liquid crystallinity, we found
that, in addition to nematic and isotropic fibril domains
\cite{Jordens}, some $\beta$-lactoglobulin fibrils were present in
circular conformations. These rings appear at the lowest interfacial density
investigated, where fibril alignment is still negligible
\cite{Jordens}, and persist in the presence of nematic fibril domains
up to high densities [see Supplementary Note 1, Supplementary Fig. S1 and S2]. Ring diameters range from $0.5-2$~$\mu\text{m}$
(Fig.~\ref{fig:AFM} and \ref{fig:PTSEM}), and are consistent whether
observed \textit{via AFM} at the air-water interface, cryogenic Scanning
Electron Microscopy (cryo-SEM) or passive probe particle tracking at
the oil-water interface, confirming that fibrils have a similar
tendency to bend at air-liquid and liquid-liquid interfaces. A small
selection of the vast variety of ring morphologies is presented in
Fig.~\ref{fig:AFM}. Highly complex structures involving several
fibrils are quite common (Fig.~\ref{fig:AFM}a, b, S1 and S2), whereas
relatively few distinct rings or tennis rackets comprise a single
fibril and can rather be thought to be intermediate assembly states $en route$ to final ring structures
(Fig.~\ref{fig:AFM}c and d) \cite{Schnurr}. Short fibrils, which could be the result of fracture due to the bending strain, exposure to air or inhomogeneous strong surface tension, also assemble into rings (Fig. S3). Alternatively, short fibrils frequently accumulate within an outer ring and align either along the circumference of this ring or parallel to each other in the
center, with minimal contact with the ring itself (Fig.~\ref{fig:AFM}b
and e).

The long-lived presence, and hence inferred stability, of these
self-organized conformations was confirmed by passive probe particle
tracking experiments performed at the oil-water interface, where
fluorescently-labelled spherical tracer particles (diameter
$\approx{774}$~nm) were observed to move in near-perfect circles or
sickle-shaped trajectories over the course of three to four minutes. A
simple pathway for ring formation could be the presence of nano- or
microbubbles at the liquid surface, which give fibrils the opportunity
to bend around their circumference \cite{Martel}. This would, however,
also lead to a distortion of the peptide layer (see Materials and
Methods) at the interface; once the sample has dried, the bubble would
have disappeared but still be visible in AFM images as a height
discontinuity through the `bubble'. The absence of such observations
in AFM (Fig. S4), or of bubbles (cavities) in the cryo-SEM images
(Fig.~\ref{fig:PTSEM}), indicates that there is an inherent
predisposition of the fibrils to bend, which then leads to circle
formation upon interaction with a liquid surface.

\subsection{Fibril Free Energy}
Understanding these data requires a study of how surface effects
influence the shape of fibrils (or indeed filaments). We consider an
inextensible fibril of length $L$, represented as a twisted ribbon with chiral wavelength $\lambda$ and pitch angle $\theta_p=\cot^{-1}(2\pi R/\lambda)$, where $R$ is the inscribing radius of the twisted ribbon (see Supplementary Note 2). We parametrize the shape by $\uvec{t}(s)$, the   direction parallel to the central axis of the ribbon, or equivalently the tangent vector of the fibril. The ribbon twists around its axis $\uvec{t}(s)$ by the angle $\phi(s)$.  We will parametrize the bending in terms of the angular rate of deflection $\dot{\boldsymbol{\Theta}}=\uvec{t}\times\dot{\hat{\textbf{t}}}$, where $\mathbf{\kappa}(s)=d\uvec{t}/ds\equiv\dot{\hat{\textbf{t}}}$ is the local curvature. Hence, $\dot{\boldsymbol{\Theta}}=\kappa \uvec{n}$, where $\uvec{n}$ is the axis about which the tangent vector is deflected during a bend. For a fibril confined to bend on a surface, we take $\uvec{n}$ to be outward surface normal vector (pointing \textit{into} the liquid), so that  $\kappa$ can be either positive or negative.  The free energy is given by  \cite{MarkSigg94b}
\begin{align}
G_{\textrm{fib}}&=\int_{0}^{L}ds  
\left\{\frac{B}{2}{\dot{\Theta}}^{2}  +
  \frac{C}{2}\left(\dot{\boldsymbol{\phi}} - \vec{q}\right)^{2}  
 + \vec{D}\cdot\dot{\hat{\textbf{t}}}\times(\dot{\boldsymbol{\phi}} 
  - \mathbf{q})\right\} \label{FullBulk}\\
&=\int_{0}^{L}ds  \left\{\frac{B}{2}\,\kappa^2 +\ldots\right\} . 
\end{align}
The first term penalizes bending, and $B$ is the  bending modulus. The second term penalizes twist relative to the native helical twist, which is parametrized by the chiral wavenumber $q=2\pi/\lambda$. Here,  $C$ is the twist modulus. The vector $\vec{D}$ represents the twist-bend couplings allowed by a polar fibril with a non-symmetric local cross section \cite{MarkSigg94b}. In this work we will focus on the bend degrees of freedom, since in filaments with free ends, such as those considered here, the  twist degrees of freedom will relax to accomodate any imposed bend.

A polar twisted fibril has an anisotropy that distinguishes `head'
from `tail' directions along the fibril axis; in F-actin this
`polarization' arises from the orientations required of G-actin
monomers to effect self-assembly \cite{Howard2001}; in an
$\alpha$-helix the N-C polymerisation breaks the polar symmetry and in
cross-$\beta$ amyloid fibrils such as those studied here the polarity
is due to the molecular packing of $\beta$-sheets
\cite{rogers2006investigating,fitzpatrick2013atomic,cohen2013proliferation}.
The polarity is reflected in variations in molecular structure along
the exposed surface of the twisted ribbon. When this structure is
placed in a heterogenous environment, as occurs near a solid surface
or when immersed within a meniscus between two fluids (or fluid and
gas), the inhomogeneity of the environment generally leads to
unbalanced torques on the body (see Supplementary Note 2, Fig. S5 and S6 for details), even when local
forces have balanced to place the fibril at the interface. A
non-symmetric body, such as a chiral and polar fibril, can thus
experience an effective spontaneous curvature
\cite{isambert1995bending}.

\begin{figure}
\begin{center}
{\includegraphics[width=0.5\textwidth]{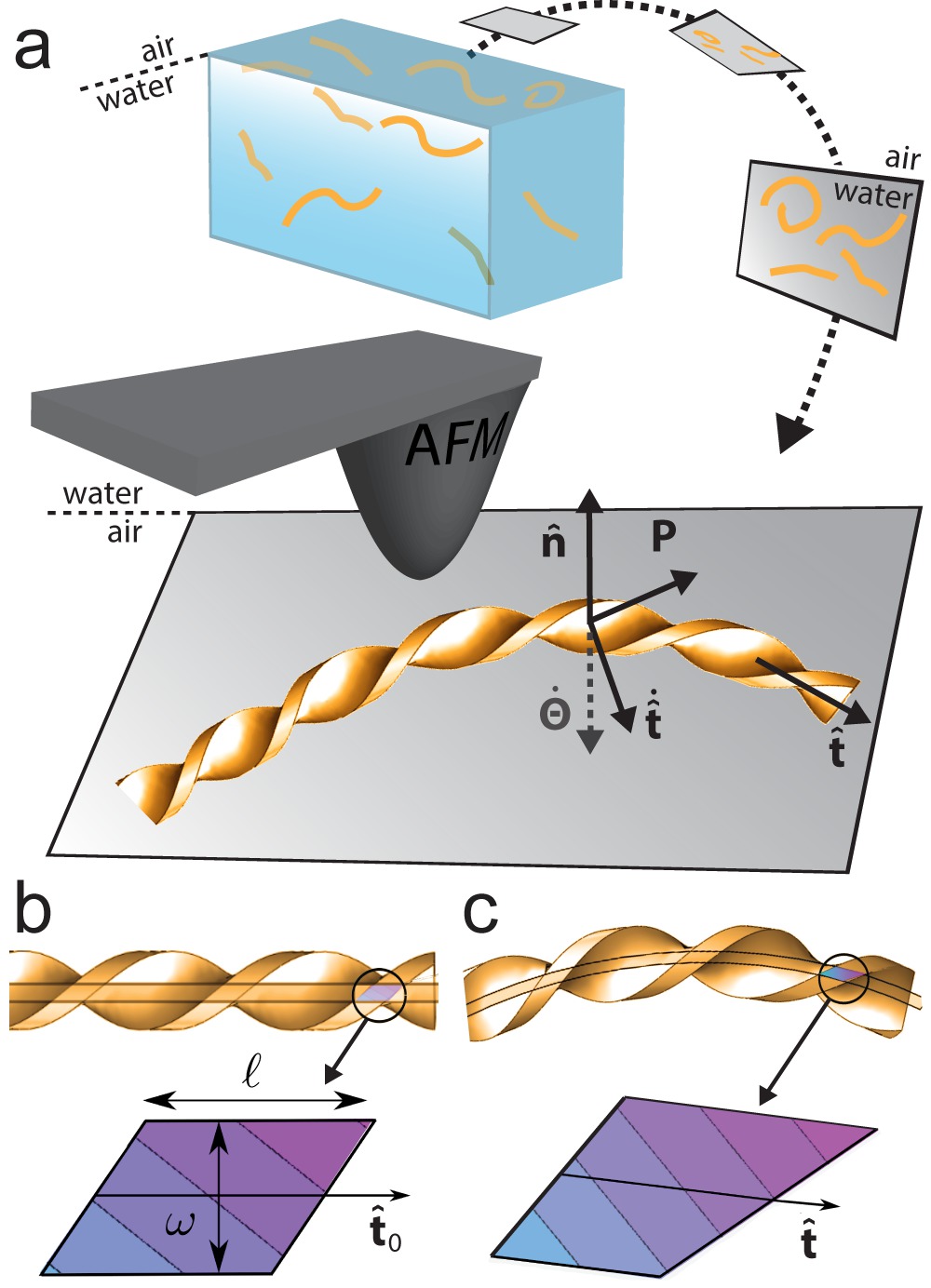}}
\end{center}
\caption{Twisted ribbon against a surface. ({\bf a}) After horizontal transfer of the interfacial fibril layer, the AFM tip probes the fibrils from the side that was originally pointing towards the water phase. ({\bf b, c}) The contact area as seen through the interface from the air-side is a
  parallelogram ({\bf b}), which deforms
  asymmetrically when the fibril is bent ({\bf c}). This leads to a greater contact area by  one  'charge' (indicated by color) of the polar interaction, which implies a preference for
  one sign of bend and thus a spontaneous curvature. The example shown
  is that of a bend that decreases the contact energy. The symmetry
  breaking of the polar region upon bending has been amplified for
  visualisation purposes.}
\label{fig:surfacecartoon}
\end{figure}

To demonstrate this effect, we consider a fibril adsorbed
\textit{onto} a planar surface with which it interacts, rather than
immersed \textit{within} a meniscus. The effects are qualitatively the
same, but the details are easier to understand in the adsorbed
case. The surface and the adsorbed ribbon interact \textit{via} numerous
molecular interactions \cite{israelachvili}. Although in principle
\textit{all} atoms in the fibril interact with every point on the
surface due to Coulomb interactions, screening limits the interaction
to only the adsorbing surface. Long-range dispersion interactions are
also irrelevant for fibrils that are induced to bend or twist within
the plane, since the change in this energy will be negligible. Hence,
we consider the following surface free energy
\begin{align}
G_{\textrm{surf}}&=\frac{2L}{\lambda} \int_{S} \left[\bar{\gamma} + \delta\gamma(\mathbf{r})\right]d^2r, 
\end{align}
where $\lambda$ is the twist pitch or wavelength, the average surface energy
$\bar{\gamma}$ controls adsorption, and $S$ is the contact area of a
the ribbon, which occurs every half wavelength. The asymmetry
$\delta\gamma(\mathbf{r})$ reflects the polar nature of the
interaction and can vary from repulsive to attractive along the repeat
patch. A polar moment (with dimensions of energy) of the interaction
can be defined by
\begin{equation}
\vec{P}=\frac{2}{\lambda}\int_{S} \,\vec{r}\,\delta\gamma(\vec{r})\,d^2r,
\end{equation}
where $S$ is the area of the patch where the fibril contacts the surface. The polar moment $\vec{P}$ is determined by the nature of the interaction with the surface, and is thus not an intrinsic property of the fibril alone. Fig.~\ref{fig:surfacecartoon} shows an example in which the
surface patch is a parallelogram with length $\ell$ and width
$\omega$. For a simple surface potential $\delta\gamma(\vec{r})=\varepsilon (x\cos\Phi+y\sin\Phi)$, where the coordinate $x$ is parallel to the fibril axis coordinate $s$, the polar moment (see Supplementary Note 2) has magnitude $P=\alpha(\theta_p,\Phi)\omega^3\ell\varepsilon/\lambda$. Here,  $\alpha(\theta_p,\Phi)$ is a geometric prefactor whose sign depends on the polarization and chirality, and parametrizes the degree to which the symmetric parellelogram is deformed into a non-symmetric shape to favor one sign of surface `charge'.

When the twisted ribbon is bent the ribbon-surface contact area
changes shape, so that either the repulsive or
attractive part of the polar interaction has more contact with the surface,
 depending on the sign of the
bend (Fig.~\ref{fig:surfacecartoon}c). This leads to a spontaneous
curvature. The contribution of bending to the overall interaction
energy can then be written as a chiral coupling between the bending rate
$\dot{\boldsymbol{\Theta}}$ and the polar moment $\vec{P}$: 
\begin{align}
G_{\textrm{surf}} &= \int_0^{L}\!\!ds\left\{ -{A}\,\dot{\boldsymbol{\Theta}}\cdot\uvec{t}\times
\vec{P}  + \ldots\right\} \label{eq:surf} \\
&= \int_0^{L}\!\!ds\left\{-{A}\,\kappa\,\uvec{n} \cdot \uvec{t}\times\vec{P}
  + \ldots\right\} . \label{eq:surf2} 
\end{align}
The vector product is the simplest term which has no mirror symmetry, and is thus appropriate for a chiral filament. Moreover, under $s\rightarrow -s$ both $\kappa$ and $\vec{t}$ change sign, whereas $\vec{P}$ does not, so that the free energy is also reparametrization-invariant. The dimensionless geometric factor ${A}$ and the moment $\vec{P}$ depend on
the details of the surface free energy $\delta\gamma(\vec{r})$
interaction potential $U$, the contact area shape, and its deformation
under bending. The polar moment $\vec{P}$ depends on the surface normal vector through its vector nature and the details of the surface-fibril interaction. The ellipses indicate other terms induced by the
surface, such as contributions to the bend-twist or curvature moduli,
or a spontaneous twist. We choose the convention that the surface
  normal vector $\uvec{n}$ points \textit{away} from the surface and thus into the fibril. An example free energy $G_{\textrm{fib,P}}$ is calculated in the Supplementary Information for a simple model contact potential.
  
The curvature in Eq.~\ref{eq:surf2} carries a sign: for $\kappa>0$ the fibril bends in a right-handed sense around the surface normal vector $\uvec{n}$, while for $\kappa<0$ the fibril bends in a left-handed sense. The  process of transferring the surface layer for AFM observation orients the surface normal towards the AFM observer, so that observation is from the liquid side towards the air side (Fig.~\ref{fig:surfacecartoon}). Consider a polarization such that 
   $\vec{P}\cdot\uvec{y}=\sin\Phi$,  where $\Phi=+\pi/4$, and choose 
   $\uvec{t}\parallel\uvec{x}$ (as observed in the AFM image; see Fig.~\ref{fig:surfacecartoon}), where $\uvec{x}\times\uvec{y}=\uvec{n}$. This
    implies $\uvec{n}\cdot{\uvec{t}}\times\uvec{P}>0$. Consider a bend as shown in Fig.~\ref{fig:surfacecartoon}, in which $\dot{\boldsymbol{\Theta}}\cdot\uvec{n}=\kappa$, where $\kappa<0$. In Supplementary Note 2 we find ${A}>0$, so that this bend ($\kappa<0$) increases the energy, and thus $\kappa>0$ is favored. Similarly, for the opposite sign of $\uvec{t}\times\vec{P}$ a negative curvature $\kappa<0$ is favored.

    The competition between the surface energy (Eq.~\ref{eq:surf2}) and
    the ordinary fibril bending energy (Eq.~\ref{FullBulk}) leads, by minimization,
    to a spontaneous curvature $\kappa_0$ given by (see SI)
\begin{equation}
\kappa_0= \frac{{A}}{B}\, \uvec{n}\cdot\uvec{t}\times\vec{P}. 
\label{eq:c0}   
\end{equation}
This is equal to $\varepsilon \omega^3 \ell\,\alpha(\theta_p,\Phi)/B\,\lambda\sin^2\theta_p$ for the simple surface potential $\delta\gamma(\vec{r})=\varepsilon (x\cos\Phi+y\sin\Phi)$.
Isambert and Maggs \cite{isambert1995bending} articulated how a surface can induce spontaneous curvature in a polar and chiral filament. They proposed a phenomenological free energy with an explicit spontaneous curvature that depends on the twist angle, and a surface interaction that breaks polar symmetry. Hence, they have actually introduced a spontaneous curvature `by hand'. Conversely, we present a model in which a polar surface interaction is itself chiral by virtue of the local chirality of the filament, and this gives rise to an effective spontaneous curvature as a result of total energy minimization. Therefore, the functional form of the resulting spontaneous curvature differs from that proposed in Ref.~[30].

Enhanced curvature is expected for amyloid fibrils with fewer
filaments (as confirmed in Fig.~\ref{fig:thickness}), which will have
smaller bending moduli $B$, or for fibrils with larger polar moments
$P $ and thus stronger surface interactions. In addition, the specific details of the surface deformation encapsulated in the function $\alpha(\theta_p,\Phi)$ play an important role: fibrils for which the deformation leads to a more symmetric contact area will have a stronger geometric factor and thus a greater expected spontaneous curvature.

\subsection{Non-Gaussian Curvature Distributions}
\begin{figure}
\begin{center}
{\includegraphics[width=0.65\textwidth]{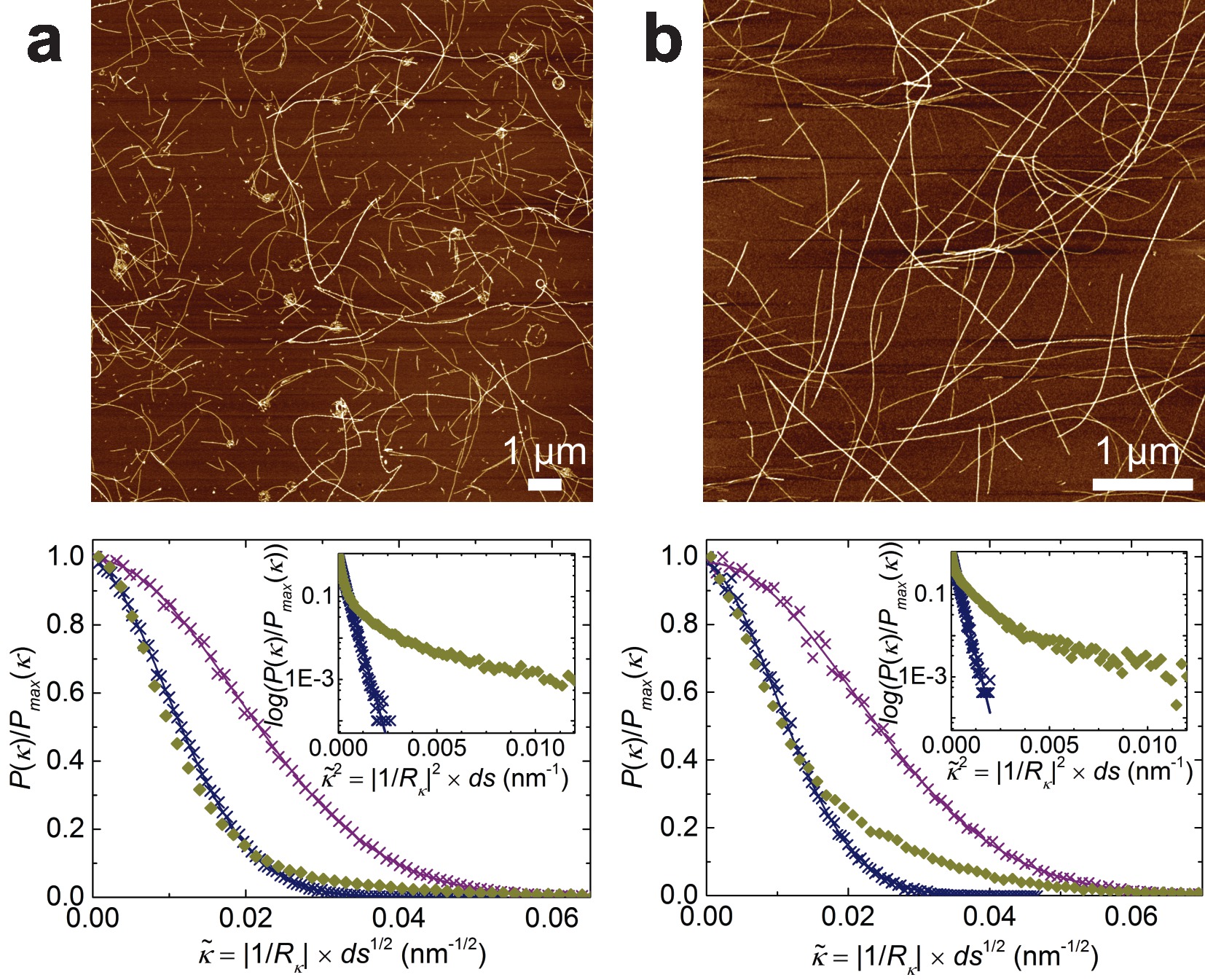}} 
\end{center}
\caption{ Fibrils exhibit a spontaneous curvature when adsorbed to a
  surface. Upper panel: zoomed in images of $\beta$-lactoglobulin
  fibrils ({\bf a}) at the air-water interface after $t=10$~minutes of
  adsorption from a $c_{\text{init}}=0.001\%$~w/w fibril suspension and ({\bf b})
  deposited onto mica for $2$~minutes from the bulk with
  $c_{\text{init}}=0.1\%$~w/w. Lower panel: probability distributions of
  normalized absolute local curvatures $\kappa$ extracted from the full ({\bf a})
  $30\times30$~$\mu\text{m}$ (see Appendix for full image) and ({\bf b})
  $5\times5$~$\mu\text{m}$ images (green diamonds) with a $ds$ of $24$
  and $9.8$~nm, respectively. The curvature distribution of simulated
  WLCs generated using all relevant parameters from the corresponding
  AFM image (see Methods) is shown as purple crosses and is
  successfully fitted with a Gaussian probability distribution
  function (purple line). Tracking these WLCs results in a change in
  the probability densities (blue crosses) but the values are still
  Gaussianly distributed (blue line). Plotting the normalized
  probabilities in logarithmic scale as a function of $\kappa^2$ clearly shows fat
  tails and thus the presence of spontaneous curvature only in real fibrils
  (Insets in the lower panel).}
\label{fig:curvature}
\end{figure}

Consider a segment of arc length $ds$ of a wormlike chain (WLC). The
probability ${\cal P}(\kappa)$ of finding this segment curved with
curvature $\kappa=1/R_\kappa$, where $R_\kappa$ is the radius of curvature, is
governed by the bending modulus and should be Gaussianly distributed,
${\cal P}(\kappa)\sim \exp\left\{-ds\,\ell_p \kappa^2/2\right\}$, where
$\ell_p=B/\kbT$ is the persistence length. Deviations from the
WLC model can be quite common, as with toroidal DNA \cite{Noy,Seaton},
in which the nucleic acids have a smaller persistence length at short
length scales \cite{Noy}. The presence of rings in our system suggests
a characteristic intrinsic curvature or length scale, in addition to
the usual $\ell_p$. For quantitative analysis, we have extracted the
$xy$ coordinates of fibrils from images acquired at low interfacial
fibril densities after short adsorption times, where interactions and
contact between fibrils are still minimal, and calculated ${\cal
  P}(\kappa)$ (Fig.~\ref{fig:curvature}; see Materials and Methods). Any
rings present on the image were excluded from the analysis, since
their closed topology would introduce an additional constraint.
To benchmark this approach, we first generate conformations based on
the discrete WLC model with the $\ell_p$ obtained from the 2D mean
squared end-to-end distance of fibrils at the air-water interface
\cite{Rivetti}. These conformations are used to create artificial
images of WLC polymers with the same resolution as the AFM images and
then subjected to the same tracking algorithm used for analyzing the
real fibril image. Fig.~\ref{fig:curvature} shows the normalized
probability distribution of curvatures ${\cal P}(\kappa)$/${\cal
  P}_{max}(\kappa)$ for both the original WLCs and the corresponding
tracked conformations (see Methods). In the tracked conformations the distribution shifts towards lower curvatures: this change is due to finite image resolution (Fig.~\ref{fig:curvature}a). Importantly, however, both distributions are Gaussian. In contrast, and as expected from the
theoretical considerations put forth above, the normalized ${\cal
  P}(\kappa)$ for real fibrils adsorbed at the air-water interface can
indeed not be fitted with a single Gaussian distribution function but
has a pronounced fat tail instead.

It has been argued that differences in $\kappa_0$ are to be expected depending on the strength of
adsorption to the surface \cite{Joanicot} and on whether the polymer
is in 3D or 2D \cite{Rappaport}. To test this, we compare the curvature distributions from fibrils adsorbed to the air-water interface and transferred horizontally to mica (Fig.~\ref{fig:curvature}a) to fibrils deposited onto mica from a drop of the bulk solution (Fig.~\ref{fig:curvature}b). The modified Langmuir-Schaefer AFM
sample preparation is a 2D to 2D transfer from a liquid onto a solid
surface, which is much faster (milliseconds) than the slower (seconds)
3D to 2D equilibration obtained by depositing onto a solid substrate
from bulk \cite{Rivetti}. The bending probability of fibrils
adsorbed from the bulk to mica, where no rings are observed, was also found to deviate
from a typical Gaussian distribution (Fig.~\ref{fig:curvature}b). Fibrils hence bend as a result of their exposure to the inhomogeneous environment of solid-liquid, liquid-liquid, and gas-liquid interfaces, independently of how they initially adsorbed at these phase boundaries.

\subsection{Average Fibril Thickness Determines Propensity to Bend}
\begin{figure*}
\begin{center}
{\includegraphics[width=0.65\textwidth]{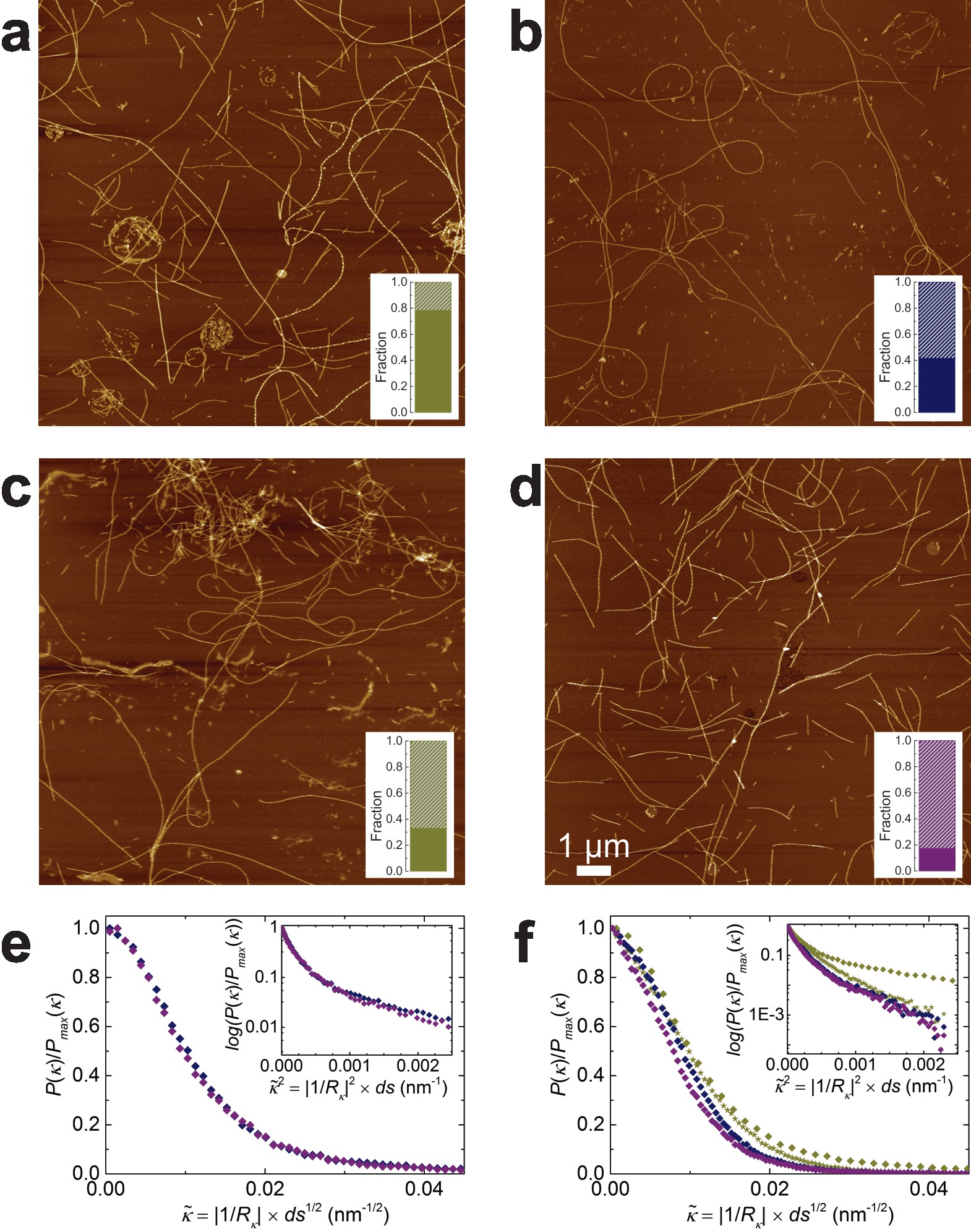}}
\end{center}
\caption{Fibril thickness affects probability of observing rings at the air-water interface. ({\bf a}-{\bf d}) Fraction of double- (solid bar) and triple-stranded fibrils (striped bar) in batches of fibrils formed from $\beta$-lactoglobulin obtained from three different suppliers. Data correspond to the ratio of the area under each of the two peaks to the total area of the average height per fibril distribution obtained on AFM samples of fibrils at the air-fibril solution interface with $c_{\text{init}}=0.001\%$~w/w. Rings are observed most frequently in batch 1 from source 1 ({\bf a}), looped structures are found in fibrils from source 2 ({\bf b}), and  batch 2 from source 1 ({\bf c}), whereas fibrils from source 3 do not form rings ({\bf d}). The scale bar applies to all AFM images. ({\bf e}) Probability distributions of normalized absolute local curvatures $\kappa$ for only double- (blue diamonds) and only triple-stranded fibrils (purple diamonds) for the sample shown in {\bf a} and Fig.~\ref{fig:curvature}{\bf a}. Less spontaneous curvature is found for thicker fibrils (inset). ({\bf f}) Probability distributions of normalized absolute local curvatures $\kappa$ for the samples shown in {\bf a} (green diamonds), {\bf b} (blue diamonds), {\bf c} (green stars) and {\bf d} (purple diamonds). Samples with higher fractions of thicker fibrils have less pronounced fat tails in their normalized ${\cal
  P}(\kappa)$ distribution (inset). }
\label{fig:thickness}
\end{figure*}
As noted above, we predict a larger fibrillar diameter to imply a
larger bending modulus, and hence a smaller likelihood of bending
spontaneously (according to Eq.\ref{eq:c0}). This was confirmed by
studying fibrils from different batches of preparation, as well as
from different suppliers.  Fig.~\ref{fig:thickness} shows ratios of double- to triple-stranded fibrils for $\beta$-lactoglobulin fibrils produced from native
protein obtained from three different suppliers. Non-identical
distributions can be expected due to different fibril processing
conditions (sample volumes, shearing and stirring histories) between
batches, and/or genetic variants between suppliers \cite{Qin}. This
then affects the individual filament thickness, and number of filaments per fibril,
due to subtle differences in proteolysis. Thicker filaments, with larger bending moduli, should have much smaller spontaneous curvatures, and not be visibly curved if thick enough.
Fig.~\ref{fig:thickness} shows the distribution of number of strands per fibril, which is proportional to thickness, as determined from the AFM images. The batch with the highest
number of rings (Fig.~\ref{fig:AFM}) contains the largest amount of double-stranded fibrils
(Fig.~\ref{fig:thickness}a). By contrast, for batches of fibrils formed
with the same protocol but from protein obtained from a different
supplier, primarily three-stranded fibrils were found, which did not assemble into rings
(Fig.~\ref{fig:thickness}d). Both a second batch of fibrils from the
first source as well as a batch from a third supplier containing a more even mix of
double- and triple-stranded fibrils yielded curved conformations (Fig.~\ref{fig:thickness}b and c). By separating the data used to calculate the normalized distribution of ${\cal
  P}(\kappa)$ presented in Fig.~\ref{fig:curvature}a into double- and triple-stranded fibrils (Fig.~\ref{fig:thickness}e), we confirm that the normalized ${\cal
  P}(\kappa)$ distribution of thick fibrils has a less pronounced fat tail and these fibrils thus bend less than their thinner counterparts. A similar trend is observed for the different batches in Fig.~\ref{fig:thickness}a-d, where a higher fraction of thicker fibrils in the sample results in less curved structures at the air-water interface and less spontaneous curvature (Fig.~\ref{fig:thickness}f).

\section{Conclusions}
We provide evidence from three different and independent experimental
techniques for the presence of complex self-assembled amyloid fibril
structures at air-water and oil-water interfaces. It has previously
been reported that fibril ends are particularly reactive, as shown in
the disruption of liposomes occurring preferentially at fibril ends
\cite{Milanesi}. Their enhanced fibrillation properties as compared to
the rest of the fibrils \cite{Tyedmers, Carulla, Knowles, Xue} in addition to possible capillary interactions \cite{Botto} may play
a role in the observed tendency of fibrils to form almost-closed
rings. The genesis of these rings and loops is explained by a
spontaneous curvature arising from the interaction of polar, chiral
and semiflexible fibrils with an interface. Because a spontaneous
curvature but no ring formation was determined in fibrils at the solid
mica-liquid interface, it can be concluded that a certain degree of
mobility at the interface supports the assembly of fibrils into such
geometries. This is in agreement with the fact that amyloid fibrils
adsorbed onto a mica surface from bulk can asymptotically reach the
expected $3/4$ exponent for a self-avoiding random walk in 2D
\cite{Lara,Usov}.  The ability of fibrils to form rings correlates with the average fibril height distribution, with loops only
observed in systems where single- and double-stranded fibrils
dominate. A shift in fibril height towards more triple-stranded
populations reduces the number of high curvature counts and thus the amount of ring structures
present. It is noteworthy, however, that a spontaneous curvature is
expected also for thicker fibrils but at lower $\kappa$ because of their
higher bending modulus, meaning that only thick fibrils which are long
enough ($L \ge 2\pi/\kappa $) will be able to form full rings.  These
findings have consequences for the understanding of how fibrils
deposit $in$~$vivo$, the morphology of plaques, biomechanical
interactions of chiral filaments with surrounding tissues, and
ultimately their effect on cells and organisms. A larger natural
dynamic analogue in the form of the circular motion of polarly
flagellated bacteria near solid surfaces has been described in the
literature \cite{Lauga} and together, these results could be seen as a
new approach for the controlled design, fabrication or improvement of
nanoswimmers and -robots.

\section{Experimental}

\subsection{Fibril Formation}
Amyloid $\beta$-lactoglobulin fibrils were prepared according to the
protocol of Jung {\it et al.} \cite{Jung}. The native, freeze-dried protein
was obtained from three different sources: Davisco, Sigma, and TU
Munich \cite{Toro-Sierra}. A $2$\%~w/w solution of purified and
dialyzed $\beta$-lactoglobulin was stirred during 5 hours at
90~$^{\circ}$C and $p\text{H}$~$2$. The resultant fibrils were then
dialyzed against $p\text{H}$~$2$ MilliQ water for 5 to 7 days to
remove unconverted proteinaceous material. There is, however, evidence
that even after complete removal of non-fibrillar material, the system
will go back to an equilibrium point where both fibrils and ''free''
peptides are present. This has been proposed for the case of
A$\beta_{1-40}$ and SH3 domain fibrils \cite{ONuallain, Carulla} and
recently for $\beta$-lactoglobulin \cite{Jordens,
  RuehsJRheol}. Another pathway for the accumulation of peptides may
be the disaggregation of fibrils upon adsorption to the air-water
interface.

\subsection{Atomic Force Microscopy}
Sample preparation and atomic force microscopy (AFM) were performed as
described previously \cite{Jordens}. All samples contained no added
salt. For the modified Langmuir-Schaefer technique, a $2$~$\mu$L aliquot
of a fibril solution of desired concentration $c_{\text{init}}$ was carefully
pipetted into a small glass vial and left to stand for time $t$. For a
given $c_{\text{init}}$ the interfacial fibril density increases with $t$ as
more fibrils adsorb to the interface. A freshly cleaved mica sheet
glued to a metal support was lowered towards the liquid surface
horizontally and retracted again immediately after a brief
contact. The mica was then dipped into ethanol ($\ge99.8$\%~v/v) to
remove any unadsorbed bulk material before drying the sample under a
weak clean air flow. Alternatively, images of fibrils in the bulk were
collected by pipetting $20$~$\mu$L of the sample onto a freshly
cleaved mica. After two minutes, the mica was gently rinsed with
MilliQ water and dried with pressurized air. Sample scanning in air
was performed on a Nanoscope VIII Multimode Scanning Probe Microscope
(Veeco Instruments) in tapping mode.

\subsection{Passive Probe Particle Tracking}
A volume of $15$~$\mu$L of a $c_{\text{init}}=0.001\%$~w/w fibril sample
seeded with $0.075\%$~w/v fluorescein isothiocyanate labelled,
positively charged silica tracer particles of diameter 
$\approx{774}$~nm, was pipetted into an epoxy resin well on a
thoroughly cleaned and plasma-treated glass coverslide. Medium chain
triglycerides were poured on top so as to create a flat oil-water
interface. The motion of tracers trapped at this interface was then
recorded on an inverted microscope (Leica DM16000B) equipped with a
$63\times1.4$~NA oil HCX PlanApo DIC objective for up to $700$ frames
at a rate of $0.374$~s. Images were analysed with standard as well as
custom-written software in IDL (ITT Visual Information Solutions)
\cite{Besseling, IsaSM, Jordens}

\subsection{Cryogenic Scanning Electron Microscopy}
Samples for freeze-fracture cryogenic Scanning Electron Microscopy
(FreSCa cryo-SEM \cite{IsaNatCommun}) were prepared by creating a flat
medium chain triglycerides (MCT)-fibril solution interface in clean,
small copper holders. The fibril solution contained the same
concentration of fluorescent tracer particle as in passive probe
particle tracking experiments and were added here for easier location
of the interface during imaging. The samples were then frozen at a
cooling rate of $30000$~Ks$^{-1}$ in a liquid propane jet freezer
(Bal-Tec/Leica JFD 030) and fractured under high vacuum at
$-140$~$^{\circ}$C (Bal-Tec/Leica BAF060). After partial freeze-drying
at $-110$~$^{\circ}$C for 3 minutes to remove ice crystals and
condensed water from the sample surfaces, they were coated with a
$2$-nm thin layer of tungsten at $-120$~$^{\circ}$C. All samples were
transferred to the precooled cryo-SEM (Zeiss Gemini 1530) under high
vacuum ($\lesssim{5\times10^{-7}}$~mbar) with an air-lock
shuttle. Imaging was performed at $-120$~$^{\circ}$C with a secondary
electron detector.

\subsection{Local Curvature Determination}
A home-built fibril tracking routine based on open active contours
\cite{Rivetti} was used to extract the fibrils' $xy$ coordinates from
AFM images with a tracking step length $\Delta s\approx1$ pixel between two
subsequent points along a tracked fibril. Any fibrils involved in ring
formation as well as those deposited from the subphase (for example
the bright ones running from top left to bottom right of the image in
Fig. S7) were discarded from the analysis. The absolute local curvature
$\kappa=|1/R_\kappa|$ with $R_\kappa$ being the radius of curvature between two
vectors \textbf{v}$_\textbf{1}$ and \textbf{v}$_\textbf{2}$ of equal
length along the fibril contour with a distance $ds$ between them, was
calculated for all fibril segment pairs in the image of interest. The curvature is given by $1/R_\kappa=($\textbf{v}$_\textbf{2}(s+ds)-$\textbf{v}$_\textbf{1}(f))/|$\textbf{v}$|ds$, where we chose  $ds=2\Delta s$. For a fibril penalized by only a bending energy, the probability of a curved segment is given by 
\begin{equation}
{\cal P}(\kappa)={\cal N}e^{-\tfrac12 \ell_p\kappa^2 ds},
\end{equation}
where ${\cal N}$ is a normalization factor, and $\ell_p=B/k_{\scriptscriptstyle B} T$ is the persisence length \cite{Doi}. The distribution  depends on the segment length $ds$ chosen for the calculation of bending.  Of course, the intrinsic persistence length is a material property and cannot depend on this discretization. Hence, the distribution of the quantity $\tilde{\kappa}=\kappa\sqrt{ds}$ is independent of the image resolution, and was used to parametrize the distribution of curvatures.

Images of WLCs were generated using the following
parameters obtained from real AFM fibril images: 
\begin{enumerate}
\item the mean and variance
of the length distribution,
\item the average fibril
radius,
\item the number of fibrils per image, 
\item $\ell_p$ determined from the fit of the
average 2D mean squared end-to-end distance 
\begin{equation}
\langle R^{2}_{2D}
\rangle=4L_c\ell_p\left[1-2\frac{\ell_p}{L_c}\left(1-e^{-L_c/2\ell_p}\right)\right],
\end{equation}
where $L_c$ is the internal contour length, 
\item  fibril tracking step $\Delta s$,
\item  and discretization $ds$.
\end{enumerate} 
The WLC coordinates from which the artifical images were created, were used as such for the calculation of ${\cal P}(\kappa)$. Additionally, the generated chains were tracked with the same algorithm used for real AFM images to illustrate the change in ${\cal P}(\kappa)$ due to resolution limits in the imaging and the apparently lower but purely Gaussian curvature distribution in tracked WLCs compared to untracked WLCs.

To calculate the curvature distribution for either double- or triple-stranded fibrils, the tracked fibril data set was separated into two based on a cut-off height obtained from the average fibril height histogram.

\begin{acknowledgement}
 Support by the Electron Microscopy of ETH Zurich (EMEZ) is
  acknowledged and the authors thank A. Schofield for the silica tracers. L. B\"oni is thanked for his help with figure design. The authors
  acknowledge financial support for S.J.  from ETH Zurich (ETHIIRA TH
  32-1), I.U. from SNF (2-77002-11), P.D.O. from an SNSF visiting
  fellowship (IZK072\_141955), and L.I. from SNSF grants PP00P2\_144646/1 and PZ00P2\_142532/1.
\end{acknowledgement}

\newpage
\appendix

\section{Appendix -- Supplementary Information}

\subsection{Persistence of Rings in the Presence of Nematic Domains}
Rings can be observed even at high interfacial fibril densities, where nematic domains cover most of the observed area as shown in Fig.~\ref{Sfig1} and \ref{Sfig2}. At this point, the rings are usually composed of many fibrils or are completely filled by short fibrils. It is worth noting, however, that some regions on the same sample can be void of rings.
There is a population of fibrils in all four batches investigated that is not consistent with the height and pitch distributions observed in \cite{Adamcik}. These fibrils are very tightly wound with a half-pitch length around $40$~nm and a maximum height between $4$ and $7$~nm and can also be seen to partake in ring formation.

\subsection{Spontaneous Bending of a Polar Twisted Ribbon at an Interface}
\subsubsection{Surface Interaction}
Most particles, including proteins, adsorb to a hydrophobic-hydrophilic interface in order to reduce the nascent hydrophobic surface tension 
\cite{Pickering1907}. In addition to this, a protein  will interact specifically with the two media according to the nature of the 
amino acids. Such interactions are both short-range (charge, hydrophobic effect, steric shapes) and long-range (dispersion interactions) 
\cite{israelachvili}. Long range interactions depend weakly on the nature of the surface, as they typically include the bulk of the 
two interface materials and the entire protein. However, the short range surface interactions depend critically on the details of the surface 
of the protein. The inhomogeneous surface of a protein results in a local moment or torque applied by the fluid at each point on the surface. 
For a helical protein immersed in a homogeneous fluid, this local torque will sum to zero across the entire surface of the protein. 
However, for a protein in an inhomogeneous environment, such as one confined to an interface, will experience a non-zero total torque $\Gamma$. 
This can induce a spontaneous curvature or twist depending on both the direction of $\Gamma$ and the strength of the intrinsic bend 
and twist moduli. 

The net torque on the protein due to its environment can be separated into contributions from short range and long range forces: 
\begin{align}
{\Gamma} &= \int_V \!d^3r\, \vec{r}\times \vec{f}_{LR}(\vec{r}) +
\Delta\int_S d^2r \,\vec{r}\times \vec{f}_{SR}(\vec{r},z)\,,\\[5.0truept]
\noalign{\noindent where $S$ and $V$ are respectively the surface and volume of the protein. The force densities are given by}\nonumber\\[-4.0truept]
\vec{f}(\vec{r}) &= - \int_{\textrm{env}}d^3r' \, 
\frac{\partial {\cal U}\left(\vec{r}-\vec{r}'\right)}{\partial (\vec{r}-\vec{r}')},
\end{align}
where the energy density  ${\cal U}(\vec{r}-\vec{r}')$ of interaction (energy per volume squared) between material in the environment at 
$\vec{r}'$ and in the protein at $\vec{r}$ can be separated into long range (\textit{e.g.} dispersion or Coulomb) and short-range (\textit{e.g.} hydrophobic or steric) interactions. Here, 
 $\Delta$ is the interaction depth within the protein (of order an amino acid in size),
and the forces 
are obtained by integrating over points $\vec{r}'$ in the environment external to the protein.  Although the net torque will generally depend on the entire shape and volume of the protein 
(because of long range dispersion and Coulomb interactions), we will illustrate the example where the effects of long range forces are 
negligible compared to those of the short range interactions. For example,  an unbalanced torque that leads to a bend in the plane of the interface will not perturb the long range energy of interaction appreciably, since there will be neligible response perpendicular to the interface.

In the case of short range interactions, we can approximate the integral over the environment as $\int_{\textrm{env}}d^3r'\simeq a\int dz'$, where the coordinate $z'$ is along the surface normal and  $a$ is the lateral area of the short interaction.  By integrating the short range potential and using the reference ${\cal U}(z=\infty)=0$, we can write the torque exerted on the surface as
\begin{align}
{\Gamma} &= a \Delta\int_S d^2r \,(\vec{r}\times\uvec{n})\,{\cal U}(\vec{r}),\\
 &\equiv\int_S d^2r \,(\vec{r}\times\uvec{n})\,\left[\bar{\gamma} + \delta\gamma\left(\vec{r}\right)\right],
\end{align}
The quantity $(a\Delta) {\cal U}(\vec{r})\equiv\bar{\gamma} + \delta\gamma(\vec{r})$ is the surface energy density of interaction introduced in Eq.~[3] of the main text.

\textbf{Fluid-fluid interface} -- At fluid-fluid interfaces an adsorbed fibril will be surrounded by both fluids, according to the  
(inhomogeneous) degree of wettability of the fibril on the two fluids. This inhomogeneous environment leads to a net uncompensated moment 
when averaged over the inhomogenous solvent environment around the fibril. Although this applies to the problem at hand, we will take the a 
pragmatic approach and illustrate the method for the simpler example of a fluid-solid interface with short-range interactions. 

\textbf{Fluid-solid interface} -- Consider a fibril adsorbed to a fluid-solid interface. Material within a short range $\Delta$, set by Coulomb 
screening, shapes of asperities, or hydrophobic effects, will interact with the solid substrate on a strip. For short range interactions a surface interaction that is symmetric from head 
to tail (a non-polar interaction) will lead to zero applied total torque, as the local torque will sum to zero, as in a homogeneous environment. However, a non-symmetric  interaction will lead 
to uncompensated torques, or bending moments, all along the length of the adsorbed fibril.

\subsubsection{Twisted Ribbon of Fixed Radius}

To make progress, we approximate the  fibril of length $L$ as a  twisted ribbon with wavelength $\lambda$, which makes contact every half wavelength with a solid 
surface on the exposed edges at the ribbon radius $R$ (Figures \ref{fig1},~\ref{fig2}).  The wavelength is related to the helical angle $\theta_p$ by 
\begin{align}
\cos{\theta_{p}}&=\frac{qR}{\sqrt{1+(qR)^2}}&
\sin{\theta_{p}}&=\frac{1}{\sqrt{1+(qR)^{2}}}, 
\label{pitch}
\end{align}
where $q=2\pi/\lambda$. The centerline of the undeformed fibril defines a tangent vector $\uvec{t}_0$, which upon bending becomes $\uvec{t}(s)$, with local curvature $\kappa=|d\uvec{t}/ds|\equiv|\dot{\hat{\textbf{t}}}|$. Equivalently, we can parametrize the curvature in terms of the vector angular rotation of the tangent vector, defined by $\dot{\boldsymbol{\Theta}}=\uvec{t}\times\dot{\hat{\textbf{t}}}$.

Rather than work in terms of torques exerted across the body, we will calculate 
the surface energy of the adsorbed fibril as a function of the fibril shape. Minimizing this energy with respect to in-plane bending will lead to an induced spontaneous curvature, which is equivalent to finding an uncompensated torque for a straight fibril. 

For a small interaction range,  $\Delta\ll R$, the interaction between the surface and the twisted ribbon can be 
approximated by the surface energy of series of strips  of thickness $ \omega=2\sqrt{2R\Delta-\Delta^2}\simeq\sqrt{8R\Delta}$ (Fig.~\ref{fig1}).  The ribbon-surface energy is given by  
\begin{align}
G_{\textrm{surf}} &= \sum_{j=1}^{2L/\lambda}\int_{S_{j}} \left[\bar{\gamma} + \delta\gamma(\vec{r})\right]d^2r\\
 &\equiv \sum_{j=1}^{2L/\lambda}G_{\textrm{polar},j},
\end{align}
where $S_{j}$ is the surface area of the $j$th interaction strip, the average surface energy $\bar{\gamma}$ represents the absorption properties of 
the ribbon, and  $\delta\gamma$ captures the polar nature of the interaction. There are $2L/\lambda$ distinct interaction strips. We assume 
that the strip has an anisotropic interaction 
potential that is polar along the direction $\uvec{u}$ within the strip, and assume the  simple form
\begin{subequations}
\begin{align}
\delta\gamma(\vec{r}) &= \varepsilon\,\vec{r}\cdot\uvec{u}\\
 &= \varepsilon\left(x\cos\Phi + y\sin\Phi\right),
\end{align}
\end{subequations}
where $\uvec{u}$ is at an angle $\Phi$ with respect to the tangent vector $\uvec{t}$.In the limit of $R\gg\Delta$ the strips can be approximated as flat, taking $\vec{r}$ as a two-dimensional vector in the plane of the surface. For short range interactions these flat strips constitute the primary interaction between the surface and the twisted ribbon.

Amyloid fibrils are composed of protofilaments, which in turn comprise layers of aligned beta sheets that are twisted about their central axis. A given fibril contains a number of protofilaments that form a ribbon, which we approximate as shown in Figure~\ref{fig2}(A). The ribbon diameter $D$ is given by the number of protofilaments in the fibrils, while the ribbon thickness  $d$  
is determined by the diameter of an individual protofilament. The ribbon length $L$ is determined by the total number of aligned beta strands.

For an undeformed fibril the interaction strip is a parallelogram tilted at an angle $\theta_p$ determined by the pitch of the ribbon, and with lengths determined by the thickness $d$ of the ribbon (the perpendicular distance between the edges) and the strip thickness $\omega$, as shown in Figure~\ref{fig2}(B). Two sides of length $\ell=d/\sin\theta_p$ are parallel to the tangent vector $\uvec{t}_0$, while the other two sides have length $ \omega/\cos\theta_p$.

When the ribbon is bent the ribbon thickness $d$ is fixed due to the fixed radius, but it curves to follow the deformed tangent vector $\uvec{t}$. Given that we are in the small bend regime, 
we approximate these sides as straight, but tilted additionally by $\bar{\phi}=\tfrac12(\phi_R+\phi_L)$ according to the average tilt of the interaction strip   (Figure~\ref{fig2}(C)).
Here $\phi_L$ and $\phi_R$ represent the additional tilts on the left and right hand sides of the interaction strip.

When the strip is bent downwards the top of the interaction strip is under tension whereas the bottom of the strip is under compression. 
Although the center of the strip is not under tension or compression, bend-stretch coupling terms may cause the ribbon to stretch or compress, leading to a new strip length $\ell'=d/\sin(\theta_p-\bar{\phi})$. This change in length contributes to the bend-stretch coupling, which is not of interest here. 

Initially, the polarity vector $\uvec{u}_0$ is at an angle $\Phi$ with respect to the tangent vector $\uvec{t}_0$. When the twisted ribbon is bent, then to first order the all vectors in the interaction strip rotate with the average rotation $\bar{\phi}$ of a particular segment; this includes both the polarity vector and the local tangent vector. However, the stretching and compression on either side of the bend cause the polarity vector to deflect non-affinely aross the strip; e.g the tilt of the polarity vector should vary smoothly between $\phi_L$  and $\phi_R$, when moving from left to right across the strip. For simplicity we will take the polarity vector to be tilted by $\bar{\phi}$ everywhere on the 
interaction strip. With this notation, the polar surface potential becomes 
\begin{equation}
\left.\delta\gamma(\vec{r})\right|_{\textrm{bent}}=\varepsilon\left[x\cos(\Phi-\bar{\phi})+ y\sin(\Phi-\bar{\phi})\right]. 
\end{equation}

\subsubsection{Polar Free Energy} 

The polar energy across a single interaction strip, or equivalently the energy per helical repeat, is then given by
\begin{align}
G_{\textrm{polar}}&=\varepsilon \int^{\tfrac12 \omega}_{-\tfrac12 \omega}\,dy\,\int^{f_R(y)}_{f_L(y)} \,dx
\left[x\cos\left(\Phi-\bar{\phi}\right)+ y\sin\left(\Phi-\bar{\phi}\right)\right], 
\\
\noalign{\noindent where} 
f_L(y)&=y\cot[\theta_{p}+\tfrac12(\phi_R-\phi_L)]-\tfrac12\ell \\
f_R(y)&=y\cot[\theta_{p}-\tfrac12(\phi_R-\phi_L)]+\tfrac12\ell \,,
\end{align}
and $\ell$ is the length of center of the  interaction strip parallel to $\uvec{t}_0$. This evaluates to
\begin{align}
G_{\textrm{polar}}=&\frac{\varepsilon\omega^3}{12}\left[\cot\left(\theta_{p}-\tfrac12\Theta\right)-\cot\left(\theta_{p}+\tfrac12\Theta\right)\right]
\left\{{\sin}\left(\Phi-\bar{\phi}\right) \right.\nonumber\\
& \left.\tfrac12{{\cos}\left(\Phi-\bar{\phi}\right)}\left[\cot\left(\theta_{p}-\tfrac12\Theta\right)+\cot\left(\theta_{p}+\tfrac12\Theta\right)\right]\right\}, 
\label{eq:polar}
\end{align}
where ${\Theta=\phi_R-\phi_L}$ is the angular deflection associated with the bend. The energy of deformation vanishes for zero bend $\Theta=0$. A positive bend $\Theta>0$ corresponds to a right hand bend, when travelling parallel to the chosen direction fo the tangent vector.

Our goal is to study the lowest order effects of the surface, which induce a spontaneous curvature 
signified by the term linear in bend $\Theta$ that arises from the small $\Theta$ approximation to
$G_{\textrm{polar}}$. The average tilt $\bar{\phi}$ can be related, geometrically, to a combination of twist and stretch, 
which leads to surface-induced bend-twist and bend-stretch couplings. 
Thus, we will expand Eq.~\ref{eq:polar} to first order in $\Theta$, and set $\bar{\phi}=0$ because we are not interested in higher order bend-twist or bend-stretch couplings (the effects of these would only be visible upon observing changes in total fibril length, or in local chirality). To lowest order in the deflection we find  
\begin{align}
G_{\textrm{polar}}&=\frac{\varepsilon \omega^3}{12\sin^2\theta_p}
(\cos\Phi\,{\cot\theta_p}+\sin\Phi)\,\Theta+\ldots\,\\
&\simeq\frac{\varepsilon \omega^3\ell}{12\sin^2\theta_p}
(\cos\Phi\,{\cot\theta_p}+\sin\Phi)\,\frac{d\Theta}{ds}.
\end{align}
In performing this expansion we have assumed that the polar direction $\uvec{u}$  (or $\Phi$) rotates affinely with the tangent; deviations from this will lead to higher 
order couplings $\Theta\,\delta\Phi$. 
Hence, the contribution to the bending energy of the entire fibril is
\begin{align}
G_{\textrm{surf}} &= \sum_{j=1}^{2L/\lambda}G_{\textrm{polar},j}\\
 &=\int_0^L\frac{2\,ds}{\lambda}\frac{\varepsilon \omega^3 \ell}{12\sin^2\theta_p}
 (\cos\Phi\,{\cot\theta_p}+\sin\Phi)\,\,\frac{d\Theta}{ds}, \label{eq:free0}
\end{align}
where we have assumed that the bend is smooth between contacts, and converted the sum to an integral via 
$\sum_j \rightarrow\int ds/\lambda$. 

The polar moment is given by 
\begin{align}
\vec{P}&=\frac{2}{\lambda}\int_{S} \,\vec{r}\, \delta\gamma(\vec{r})\,\,d^2r,\\
&=\frac{2\varepsilon}{\lambda}\int_{S} \,\vec{r}\, (\vec{r}\cdot\uvec{u})\,\,d^2r,\\
&=\frac{\varepsilon}{\lambda}\frac{\partial}{\partial\uvec{u}} \int^{\frac{\omega}{2}}_{-\frac{\omega}{2}}dy\int^{y\cot\theta_{p}+\tfrac{\ell}{2}}_{y\cot\theta_{p}-\tfrac{\ell}{2}}
 \left[x\cos\Phi+ y\sin\Phi\right]^2\, dx\\\,
&=\frac{\varepsilon \omega^3\ell }{6\lambda}
\left\{\left[\cos\Phi\left(\cot^2\theta_p + \left(\frac{\ell}{\omega}\right)^2\right) + \sin\Phi\cot\theta_p \right]\uvec{t} + \left(\sin\Phi + \cos\Phi\cot\theta_p\right)\uvec{n}\times\uvec{t}\right\}.
\end{align}
One component of $\vec{P}$ is parallel to the fibril direction $\uvec{t}$, while the other direction is perpendicular to $\uvec{t}$ and in the plane specified by normal vector 
$\uvec{n}$. Note that $\{\uvec{t},\uvec{n}\times\uvec{t},\uvec{n}\}$ form an orthonormal basis. Hence, 
\begin{equation}
\vec{P} = P_{\parallel}\vec{t} + P_{\perp}\uvec{n}\times\uvec{t},
\end{equation}
where
\begin{subequations}
\begin{align}
P_{\parallel} &= \frac{\varepsilon \omega^3\ell }{6\lambda}
\left[\cos\Phi\left(\cot^2\theta_p + \left(\frac{\ell}{\omega}\right)^2\right) + \sin\Phi\cot\theta_p \right]\\
P_{\perp} &= \frac{\varepsilon \omega^3\ell }{6\lambda}  \left(\sin\Phi + \cos\Phi\cot\theta_p\right).
\end{align}
\end{subequations}

Comparing the definition of $\vec{P}$ with the free energy $G_{\textrm{surf}}$, we can rewrite the surface energy as
\begin{align}
G_{\textrm{surf}}  &=\frac{1}{\sin^2\theta_p}\int_0^Lds\,{P_{\perp}}\,\frac{d\Theta}{ds}. 
\end{align}
In vector form, the angular rotation is given by $\dot{\boldsymbol{\Theta}}=-\uvec{n}\frac{d\Theta}{ds}$ (Fig.~\ref{fig2}), while the component $P_{\perp}$ can be extracted via $P_{\perp}=\uvec{n}\cdot\uvec{t}\times\vec{P}$. Thus, the free energy becomes
\begin{align}
G_{\textrm{surf}}  &=-\frac{1}{\sin^2\theta_p}\int_0^Lds\,\dot{\boldsymbol{\Theta}}\cdot\uvec{t}\times\vec{P},
\end{align}
which corresponds to the free energy of Equations~5-6 in the main text, with $A=1/\sin^2\theta_p$.

\subsubsection{Induced Curvature}

The total bending free energy is given by the sum of the standard bending energy and the coupling to the surface: 
\begin{align}
G_{bend} &= \int ds\left[\frac12 B \dot{\boldsymbol{\Theta}}^2 
- \frac{1}{\sin^2\theta_p}\dot{\boldsymbol{\Theta}}\cdot\uvec{t}\times\vec{P} \right] \\
&= \int ds \left[ \frac12 B \kappa^2 -
\frac{\varepsilon \omega^3 \ell}{6\lambda\sin^2\theta_p}
 (\cos\Phi\,{\cot\theta_p}+\sin\Phi)\kappa \right],
\end{align}
where the (signed) curvature is defined by $\dot{\Theta}=\kappa\uvec{n}$. The bending modulus generally includes contributions from the surface, which can be calculated based on the formalism here. However, since our intent is to demonstrate the significance of the induced curvature, we do not consider such perturbations. Moreover, the main contribution to bending is usually from internal degrees of freedom that are only weakly influenced by the surface. An exception occurs for highly charged filaments. In such cases the reduction in the dielectric constant and lack of screening near a hydrophobic surface will increase the electrostatic contribution to   $B$. 

 This bend energy is minimized by the following spontaneous curvature $\kappa_0$:
\begin{align}
\kappa_0 & = \frac{\varepsilon \omega^3 \ell}{6\lambda\sin^2\theta_p\,B}
 (\cos\Phi\,{\cot\theta_p}+\sin\Phi) \label{eq:spont}\\
 &= \frac{\varepsilon \omega^3 \ell}{\lambda\sin^2\theta_p\,B}
\alpha(\theta_p,\Phi),
\end{align}
where $ \alpha(\theta_p,\Phi)\equiv(\cos\Phi\,{\cot\theta_p}+\sin\Phi)/6$.

The sign of the induced curvature can be understood as follows. Consider $\varepsilon>0$,  a helix with an opening angle of $\theta_p=\pi/4$, and a polarization direction specified by  $\Phi=\pi/6$ (roughly as in Figs.~\ref{fig1}, \ref{fig2}). In this case there is a higher energy for exposing the upper right part of the parallelogram in Fig.~\ref{fig2} to the surface. Hence the preferred bending direction should be `up' in Fig.~\ref{fig1} (rather than the downward shown), to allow the relatively less of the costly part of the surface interaction to attain more contact with the surface. This  corresponds to a positive bend around $\uvec{n}$, given by $\dot{\boldsymbol{\Theta}}=\kappa_0\uvec{n}$ with $\kappa_0>0$ and matches the prediction in Eq.~(\ref{eq:spont}).

\pagebreak

\renewcommand{\thefigure}{S\arabic{figure}}

\begin{figure}
\begin{center}
\centerline{\includegraphics[width=1\textwidth]{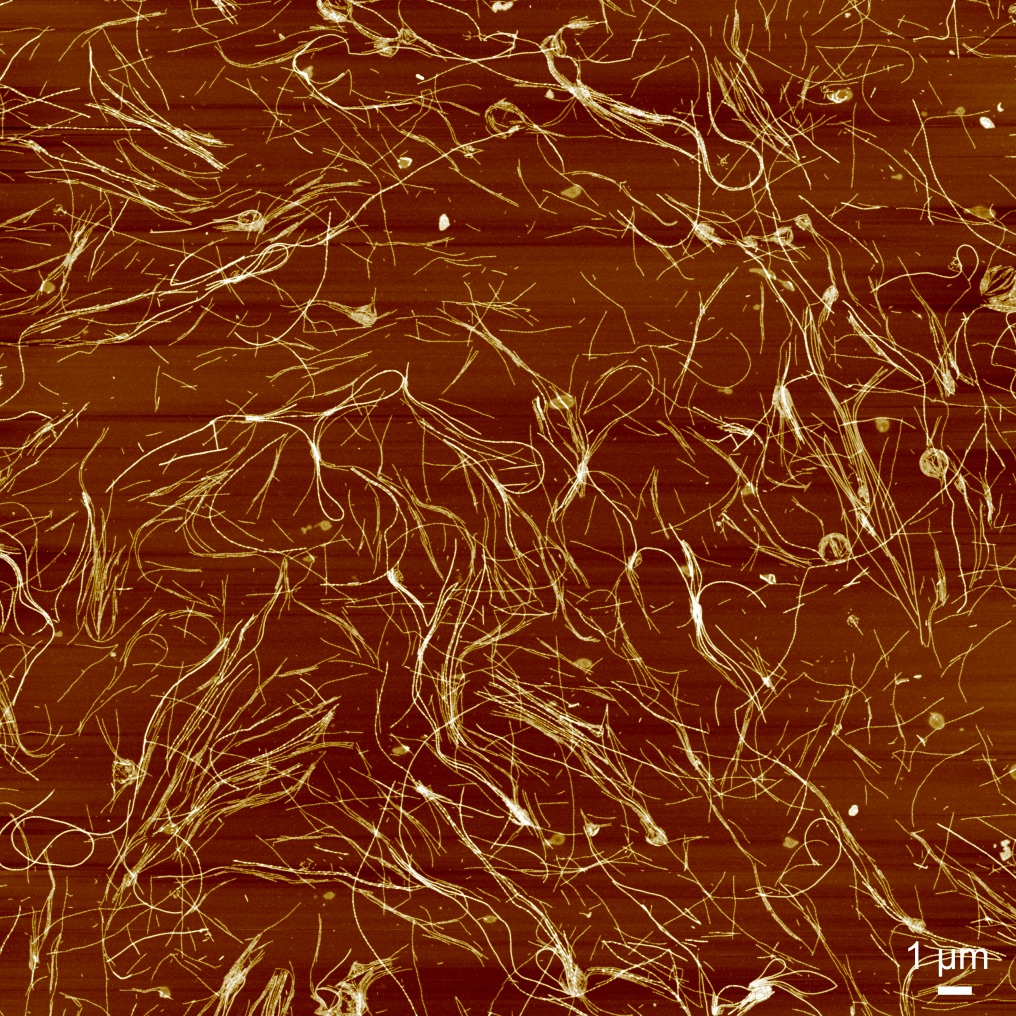}}
\caption{\label{Sfig1} AFM image of fibrils at the air-water interface after $t$=60 minutes adsorption time from a $c_{\text{init}}=0.001$\%~w/w fibril suspension.}
\end{center}
\end{figure}

\begin{figure}
\begin{center}
\centerline{\includegraphics[width=1\textwidth]{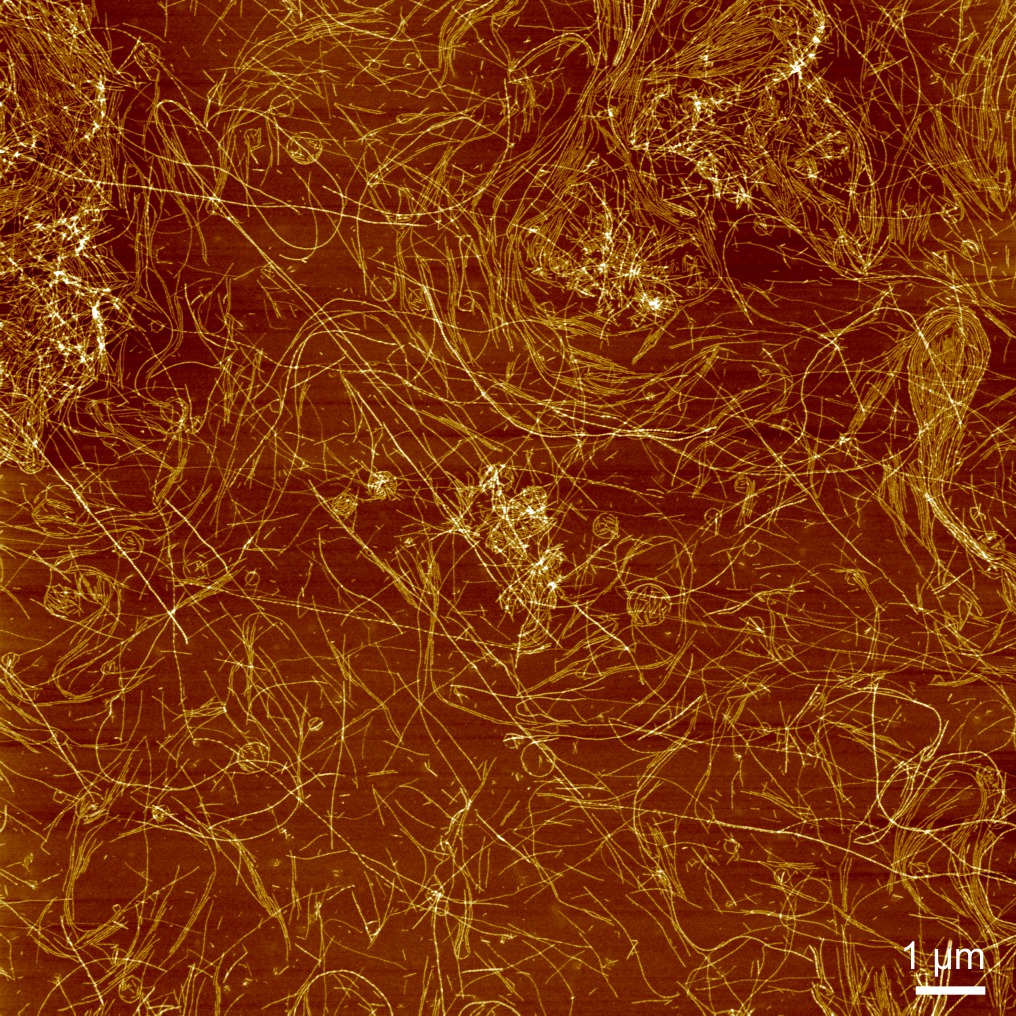}}
\caption{\label{Sfig2} AFM image of fibrils at the air-water interface after $t$=10 minutes adsorption time from a $c_{\text{init}}\approx{0.008}$\%~w/w fibril suspension. Rings coexist with nematic fibril domains.}
\end{center}
\end{figure}

\begin{figure}
\begin{center}
\centerline{\includegraphics[width=1\textwidth]{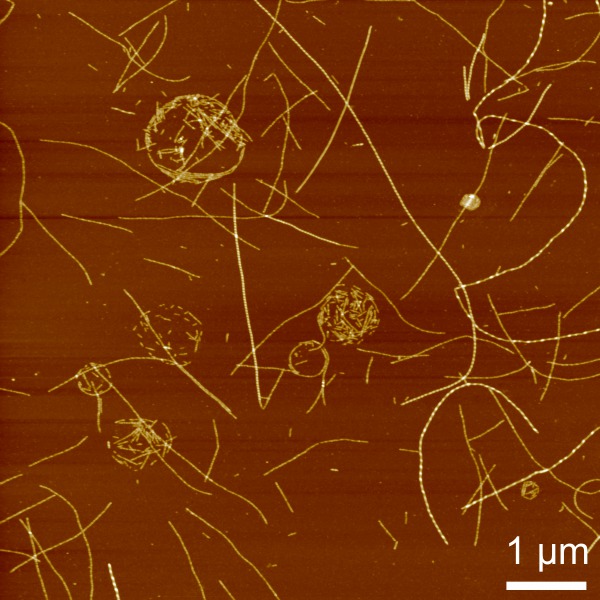}}
\caption{\label{Sfig4} AFM image of fibrils at the air-water interface after $t$=60 minutes adsorption time from a $c_{\text{init}}=0.001$\%~w/w fibril suspension. Rings are often composed of many short fibrils.}
\end{center}
\end{figure}

\begin{figure}
\begin{center}
\centerline{\includegraphics[width=1\textwidth]{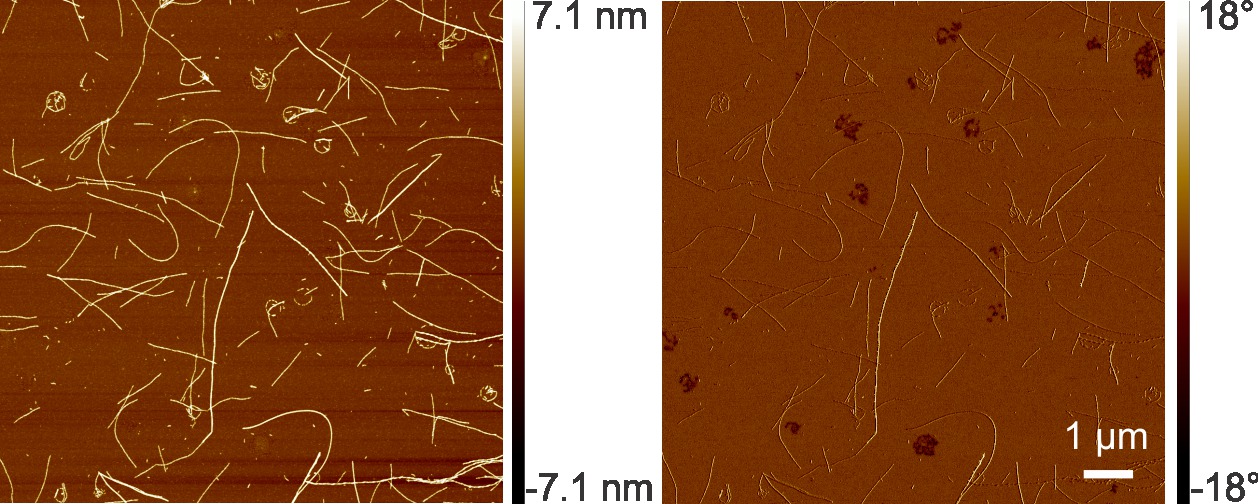}}
\caption{\label{Sfig5} AFM height and phase images of fibrils at the air-water interface immediately after sample preparation of a $c_{\text{init}}=0.001$\%~w/w fibril suspension. The scale bar applies to both images. Distortions in the background peptide layer are readily visible in the phase image but are rarely spherical and do not coincide spatially with fibril rings.}
\end{center}
\end{figure}

\begin{figure}
\begin{tabular}{c}
{\includegraphics[width=15cm]{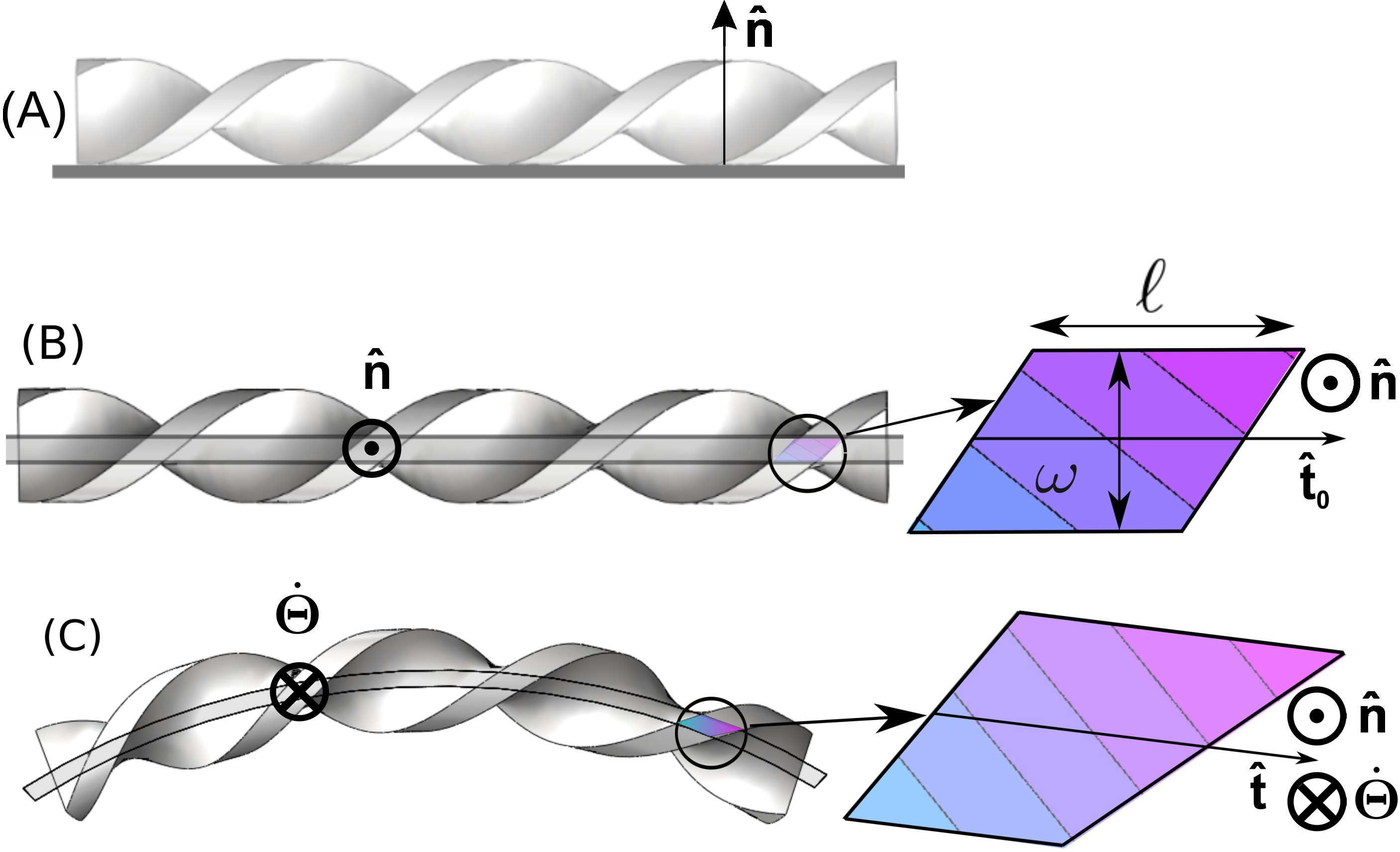}}\\[5.0truept]
\end{tabular}
\caption{(A) Helical fibril against a surface. (B) The contact area, or interaction strip, 
is a parallelogram that deforms asymmetrically (C) when the 
fibril is bent. This leads to an excess contact area by one `charge' of the polar interaction, 
leading to a preference for one sign of bend and 
thus a spontaneous curvature. A positive red `charge' and a negative `blue' charge corresponds 
to a polarization potential $\delta\gamma=\varepsilon(\cos\pi/6 + y \sin\pi/6)$, with 
$\varepsilon>0$. In this case the bend shown in (C) costs energy, and the preferred 
spontaneous curvature instead correponds to a bend $\frac{d\boldsymbol{\Theta}}{ds}=\dot{\boldsymbol{\Theta}}=\uvec{t}\times\dot{\hat{\bf{t}}}$ 
which is parallel to $\uvec{n}$.}
\label{fig1}
\end{figure}

\begin{figure}
\begin{center}
{\includegraphics[width=15cm]{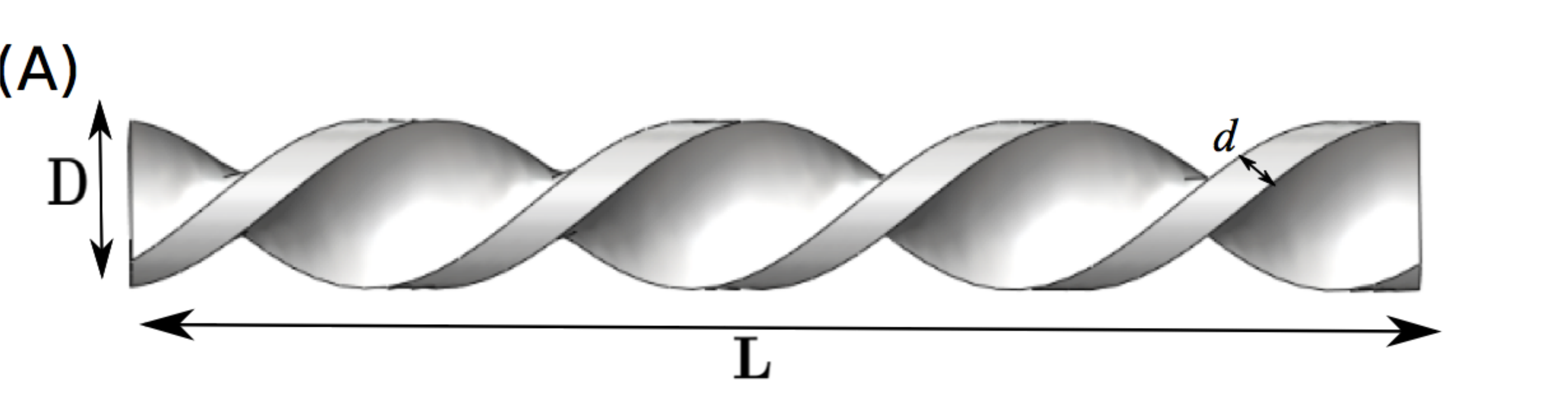}}\\
{\includegraphics[width=7cm]{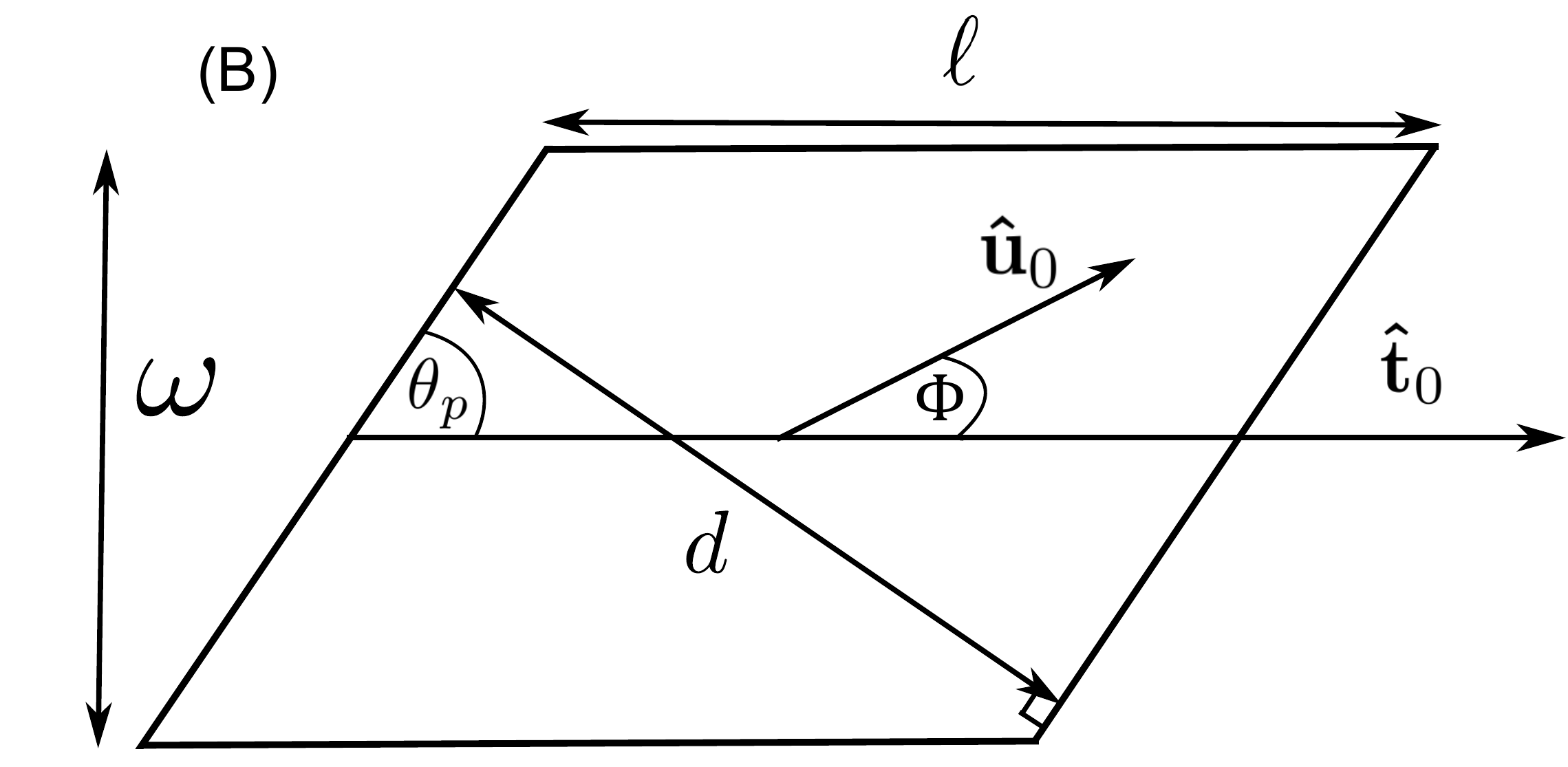}}{\includegraphics[width=8cm]{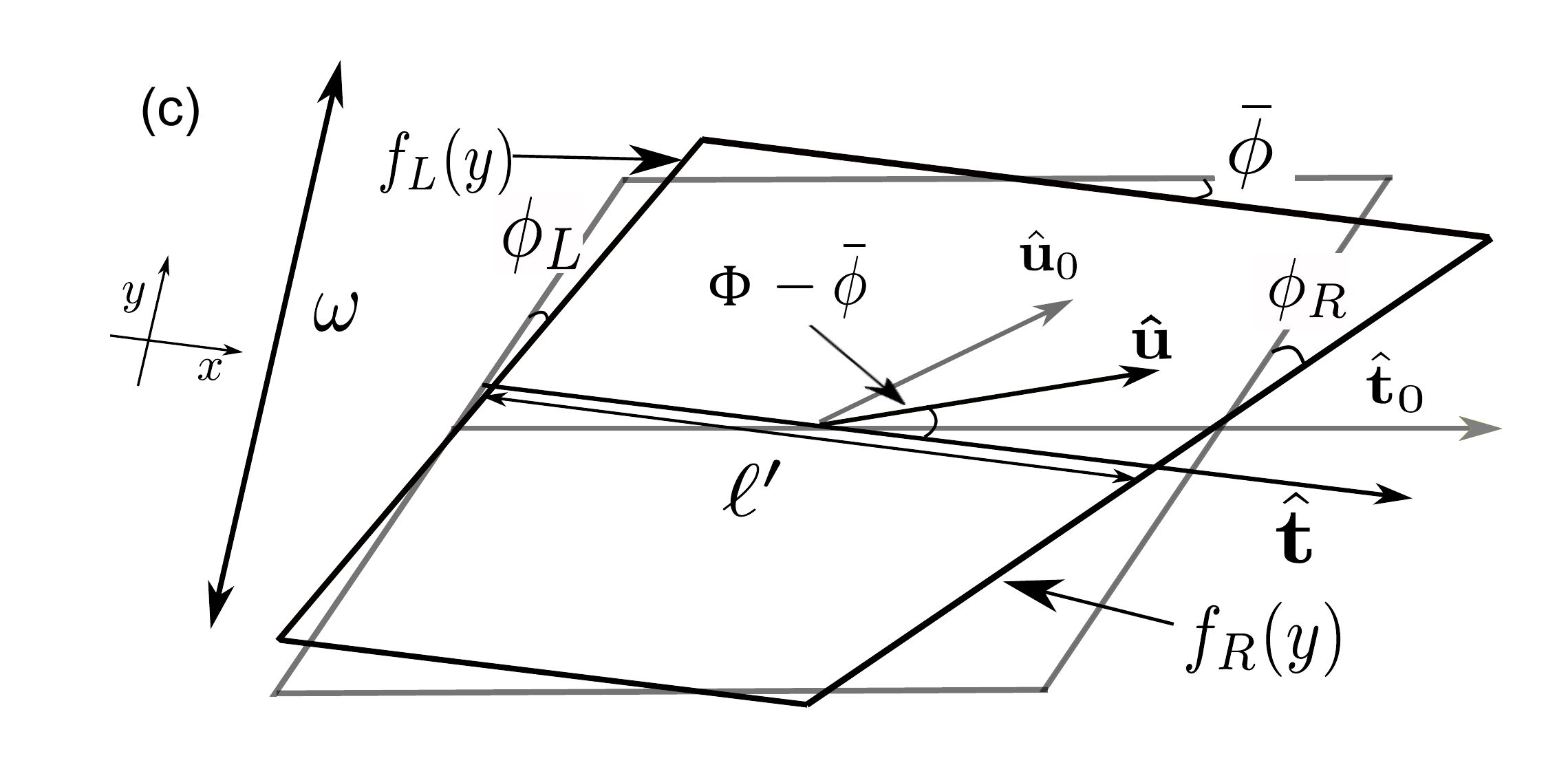}}
\end{center}
\caption{(A) Geometry of twisted ribbon.  Initial  (B) and deformed  (C) interaction strips, obtained by bending the fibril. 
The undeformed strip is shown in grey under the deformed strip. The bend causes a tilt in the two sides (right and left) depending on the change in the tangent
vector across the strip, while the top and bottom sides remain parallel to each other, but rotate with respect to the undeformed strip by $\bar{\phi}=\tfrac12(\phi_L+\phi_R)$, which describes the average tilt of the individual strip.}
\label{fig2}
\end{figure}

\begin{figure}
\begin{center}
\centerline{\includegraphics[width=1\textwidth]{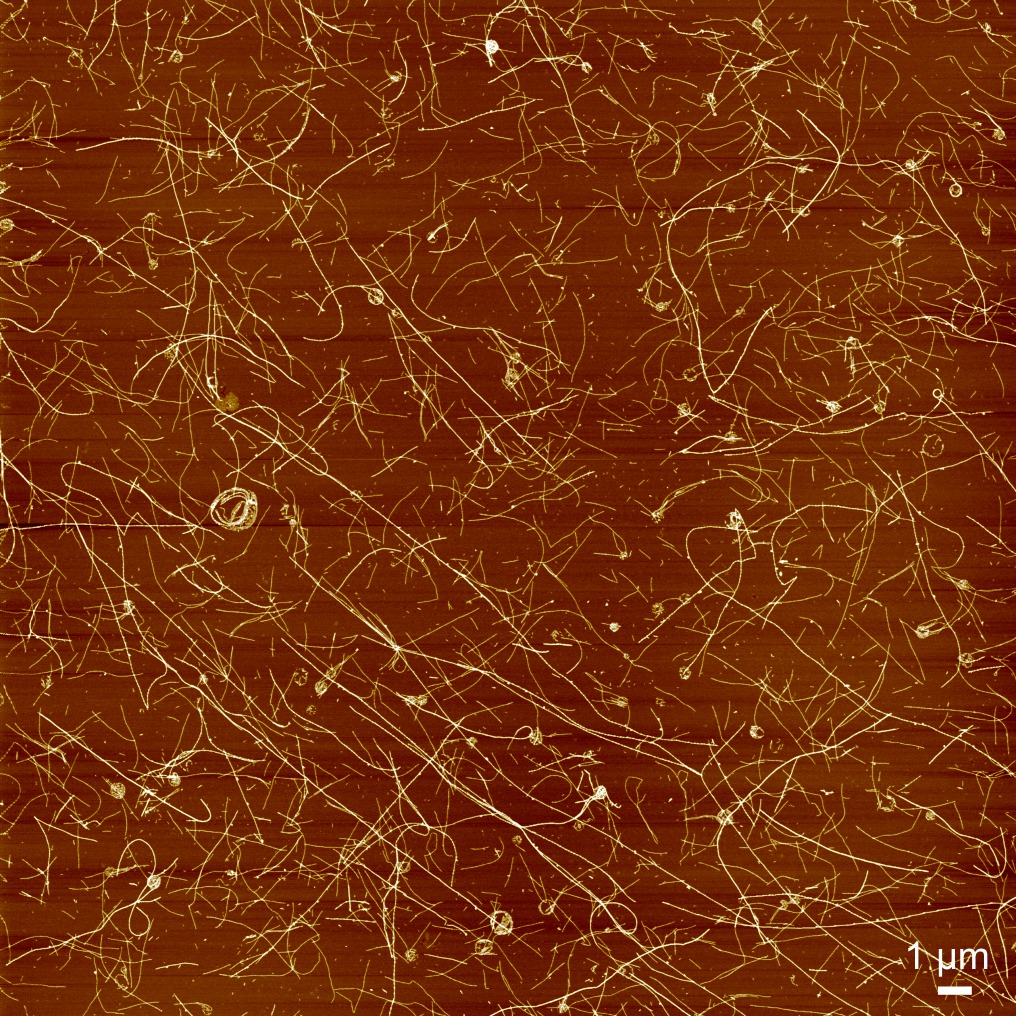}}
\caption{\label{Sfig3} Whole AFM image used for the curvature distribution analysis showing fibrils at the air-water interface after $t$=10 minutes adsorption time from a $c_{\text{init}}=0.001$\%~w/w fibril suspension.}
\end{center}
\end{figure}

\end{document}